\documentclass[aps,twocolumn,superscriptaddress, prd]{revtex4-1}

\usepackage[]{mathtools}
\usepackage{bm}
\usepackage{graphicx}
\usepackage{subfigure} 
\usepackage{float}
\usepackage[colorlinks,linkcolor=blue,anchorcolor=blue,citecolor=blue]{hyperref} 
\usepackage{verbatim} 
\usepackage{ulem} 

\begin{document}


\title{Joint Observations of Space-based Gravitational-wave Detectors: Source Localization and Implication for Parity-violating Gravity}


\author{Qian Hu}
\email[]{hq2017@mail.ustc.edu.cn}
\author{Mingzheng Li}
\email[]{sunset@mail.ustc.edu.cn}
\author{Rui Niu}
\email[]{nrui@mail.ustc.edu.cn}
\author{Wen Zhao}
\email[]{wzhao7@ustc.edu.cn}
    \affiliation{CAS Key Laboratory for Research in Galaxies and Cosmology, Department of Astronomy, University of Science and Technology of China, Hefei 230026, China}
    \affiliation{School of Astronomy and Space Sciences, University of Science and Technology of China, Hefei, 230026, China}

\date{\today}

\begin{abstract}

Space-based gravitational-wave (GW) detectors, including LISA, Taiji and TianQin, are able to detect mHz GW signals produced by mergers of supermassive black hole binaries, which opens a new window for GW astronomy. In this article, we numerically estimate the potential capabilities of the future networks of multiple space-based detectors using Bayesian analysis. We modify the public package Bilby and employ the sampler PyMultiNest to analyze the simulated data of the space-based detector networks, and investigate their abilities for source localization and testing the parity symmetry of gravity. In comparison with the case of an individual detector, we find detector networks can significantly improve the source localization. While for constraining the parity symmetry of gravity, we find that detector networks and an individual detector follow the similar constraints on the parity-violating energy scale $M_{\rm PV}$. Similar analysis can be applied to other potential observations of various space-based GW detectors.

\end{abstract}


\maketitle

\section{Introduction \label{Sec1}}

While ground-based gravitational-wave (GW) detectors are giving decent probes of high-frequency GWs~\cite{O1O2}, low-frequency GW detection still remains blank. Several proposed space-based GW detectors with frequency bands around millihertz, aiming at sources including Super-massive Black Hole Binaries (SMBHBs), Extreme Mass Ratio Inspirals (EMRIs), etc, are going to launch in early 2030s~\cite{Hu2018,Amaro-Seoane2017,Liu2020a}. Their individual properties were well studied in previous works, but space-based detector network is still a largely under-explored domain. Moreover, limited by complex response of space-based GW detectors and accompanying computation burden, most works on space-based GW detectors are based on Fisher information matrix analysis, which can only give a rough estimation of the parameter uncertainties for the potential observations, if the signal-to-noise ratio of GW detection is high enough. In this paper, we investigate the capabilities of space-based detector networks with a full Bayesian analysis. We choose two aspects to illustrate capabilities of detector networks: source localization and constraints on parity-violating (PV) gravity. 

GW source localization is a crucial step in multi-messenger astronomy, since the follow-up electromagnetic observations need the guide from GW detection. For ground-based detectors, although rapid sky reconstruction algorithm is used in online searching~\cite{Singer2016}, full Bayesian analysis is still required for further study due to its rigor and reliability~\cite{GW170817}. Recent work with Fisher information matrix analysis has shown that a LISA-Taiji network could achieve the significant improvement compared with a single detector~\cite{Ruan2019}. In this work, we study the localization improvement of detector networks LISA-Taiji and LISA-TianQin with a rigorous Bayesian framework as a complement and verification to the previous works. 

In addition to multi-messenger astronomy, GW detection also opens a brand new window for testing various theories of gravity. With the progress in both theoretical and observational researches, Einstein's general relativity (GR) is facing difficulties, such as quantization, dark matter and dark energy problems. Therefore, testing GR is still an important topic in physical research. Detectable GWs are often produced by the densest objects with extremely high-energy processes (e.g. the coalescence of binary black holes), and have weak interactions with matter during propagation~\cite{GW1,GW2}. Thus, GWs could carry strong and clean information from those extreme processes, and provide an excellent opportunity to test the gravitational theories. Space-based GW detectors are expected to detect gravitational radiations from SMBHBs, which are significantly different from current stellar-mass binary black holes. Hence, it is worthwhile to study the probability of testing gravity theories with space-based GW detectors. In this work, as an example of application, we will investigate this issue from the perspective of parity symmetry of gravity.

Parity symmetry is an important concept in modern physics. It implies the flip in the sign of spatial coordinates does not change physical laws. Since people have discovered that weak interaction is not symmetric under parity~\cite{Yang1956}, tests of parity symmetry for other interactions become meaningful and necessary. As for gravity, parity is conserved in GR, but some PV gravitational theories were proposed for different motivations. For example, in string theory and loop quantum gravity, the parity violation in the high-energy regime is inevitable~\cite{Alexander_2009,CAMPBELL1991778,Campbell_1993}. GWs probe physics in the highest energy scale, so it is nature to test parity symmetry with GWs. Parity asymmetry in gravity leads to birefringence in gravitational waves~\cite{Alexander_2009,Zhao_PVwaveform,Qiao2019,wang2013,zhu2013,qiao2020}: left- and right-hand modes of GW evolve differently in the universe. Two kinds to birefringence, amplitude birefringence and velocity birefringence, and their impact on GW waveforms, are well studied in previous works~\cite{Zhao_PVwaveform,Qiao2019}, which makes it possible to probe asymmetry in gravity.  The analysis has been applied in the current GW events, detected by LIGO \& Virgo Collaborations ~\cite{Wang2020}. In this article, we extend this Bayesian analysis to the space-based GW detection by simulating the future GW signals produced by the mergers of SMBHBs. 
We analysis simulated data and obtain the potential constraints of parity asymmetry in gravity provided by the future space-based detectors. We find that lower bound of parity-violating energy scale $M_{\mathrm{PV}}$ can be limited to $\mathcal{O}(1)$~eV by the effect of velocity birefringence and $\mathcal{O}(10^{-15})$~eV by that of amplitude birefringence.

This paper is organized as follows. In Sec.~\ref{Sec2} we give a brief introduction of parity-violating gravity, especially the GW waveform modifications. In Sec.~\ref{Sec3} the configuration and response of space-based gravitational-wave detectors are presented. Our method of parameter estimation is shown in Sec.~\ref{Sec4} and results are given in Sec.~\ref{Sec5} (localization) and \ref{Sec6} (PV gravity). In Sec.~\ref{Sec7}, we summarize our methodology and conclusions. Throughout this paper, we set $c=\hbar=1$.


\section{Parity-violating Gravity \label{Sec2}}

Parity-violating gravitational theories are well-studied in previous works~\cite{Kosteleck__2016,Yoshida_2018,Yagi_2018,Alexander_2018,Zhao_PVwaveform,silva2020astrophysical,Shaolj_2020}. In this section, we briefly summarize the results of Ref.~\cite{Zhao_PVwaveform} that gives GW waveform with PV modification. Considering a general parity-violating gravitational theory, the action takes the form
\begin{equation}
    S = \frac{1}{16\pi G}\int d^4 x \sqrt{g} (L_{\rm GR}+L_{\mathrm{PV}}+L_{\mathrm{others}}) ,
\end{equation}
where $L_{\rm GR}$ is the Einstein-Hilbert Lagrangian density $R$. $L_{\mathrm{PV}}$ is the PV term, which is determined by the gravitational theories. $L_{\mathrm{others}}$ represents the Lagrangian density of the other matters, the scalar field and the modification terms of gravity, which are not relevant to parity violation. In the flat Friedmann-Robertson-Walker (FRW) universe, GW is tensorial perturbation of the metric. We denote spatial perturbation as $h_{ij}$, which satisfies the transverse and traceless gauge, i.e. $\delta^{i j} h_{i j}=0$ and $\partial_{i} h^{i j}=0$. $h_{ij}$ can be determined by the tensor quadratic action, which reads~\cite{PhysRevLett113231301},
\begin{widetext}
    \begin{equation}
        S^{(2)}=\frac{1}{16 \pi G} \int d t d^{3} x a^{3}\left[\frac{1}{4} \dot{h}_{i j}^{2}-\frac{1}{4 a^{2}}\left(\partial_{k} h_{i j}\right)^{2}
        +\frac{1}{4}\left(\frac{c_{1}}{a M_{\mathrm{PV}}} \epsilon^{i j k} \dot{h}_{i l} \partial_{j} \dot{h}_{k l}+\frac{c_{2}}{a^{3} M_{\mathrm{PV}}} \epsilon^{i j k} \partial^{2} h_{i l} \partial_{j} h_{k l}\right)\right],
    \end{equation} 
\end{widetext}
where $a=a(\tau)$ is the conformal scale factor and $\tau$ is conformal time. A \textit{dot} means derivative with respect to the cosmic time $t$, which obeys the relation $dt = ad\tau$. $c_1$ and $c_2$ are dimensionless coefficients, which are functions of cosmic time in general. $M_{\mathrm{PV}}$ is the parity-violating energy scale, above which parity symmetry of gravity is broken. Equation of motion of the GW can be derived as follows:
\begin{equation}
    h_{\mathrm{A}}^{\prime \prime}+\left(2+\nu_{\mathrm{A}}\right) \mathcal{H} h_{\mathrm{A}}^{\prime}+\left(1+\mu_{\mathrm{A}}\right) k^{2} h_{\mathrm{A}}=0,
\end{equation}
where $A=\{R,L\}$ represents right- and left- modes, respectively. $k$ is wave-number, $\mathcal{H}\equiv a'/a$ is the conformal Hubble parameter. Throughout this paper, \textit{prime} denotes the derivative with respect to the conformal time $\tau$. The terms $\nu_{\mathrm{A}}$ and $\mu_{\mathrm{A}}$ represent modifications caused by the PV terms in Lagrangian. In the general PV gravity, they take the forms
\begin{equation}
    \begin{aligned}
        \nu_{\mathrm{A}} &=\left[\rho_{\mathrm{A}} \alpha_{\nu}(\tau)\left(k / a M_{\mathrm{PV}}\right)\right]^{\prime} / \mathcal{H}, \\
        \mu_{\mathrm{A}} &=\rho_{\mathrm{A}} \alpha_{\mu}(\tau)\left(k / a M_{\mathrm{PV}}\right).
    \end{aligned}
\end{equation}
Here, $\rho_{R} = 1$ and $\rho_{L} = -1$. $\alpha_{\nu}=-c_1$ and $\alpha_{\mu}=c_1 - c_2$ are two functions that can be determined in a specific model of modified gravity. In the specific models, $\alpha_{\nu}$ and $\alpha_{\mu}$ are functions of time through their dependence on scalar field $\phi$, which always acts as dark energy to explain the cosmic acceleration.  From cosmological observations, dark energy should be close to the cosmological constant in the late universe, which indicates that the evolution of $\phi$ is small.  Therefore, we can approximately treat them as constants in our calculation. In this work, we consider they are $\sim \mathcal{O}(1)$ by absorbing them into $M_{\mathrm{PV}}$. Difference in equation of motion of two circular polarization modes leads to parity asymmetry in GWs, that is to say, right- and left-hand modes have different behaviors during propagation, which is called birefringence. It has been proved that $\nu_{\mathrm{A}}$ leads to different damping rates of two polarizations in propagation, which induces the different amplitudes of GW signals. $\mu_{\mathrm{A}}$ modifies the dispersion relations of GWs, hence two polarizations have different velocities. Phenomena mentioned above are called amplitude birefringence and velocity birefringence respectively.

Birefringence in PV gravity induces phase and amplitude modifications in GW waveform. In general, GW waveform of PV gravity in frequency domain can be expressed as
\begin{equation}
    h_{\mathrm{A}}^{\mathrm{PV}}(f)=h_{\mathrm{A}}^{\mathrm{GR}}(f)\left(1+\rho_{\mathrm{A}} \delta h\right) e^{i \rho_{\mathrm{A}} \delta \Psi},
\end{equation}
where 
\begin{equation}
    \label{proptof}
    \begin{aligned}
        \delta h(f) &= - A_\nu \pi f, \\
        \delta \Psi(f) &= A_\mu (\pi f)^2 / H_0
    \end{aligned}
\end{equation}
are amplitude and phase modifications. Generally, both of them exist in PV gravity. Note that $\delta \Psi(f)$ is about 20 orders lager than $\delta h(f)$~\cite{Zhao_PVwaveform}, it is reasonable to only take $\delta \Psi(f)$ into consideration when considering PV effects. However, in some special cases, say, Chern-Simons gravity~\cite{Alexander_2009,CAMPBELL1991778}, $\delta h(f)$ exists while $\delta \Psi(f)=0$. Therefore, it is also necessary to constrain the amplitude modification. In this work, for simplicity, we only discuss PV GW waveform with only phase modification or amplitude modification. The former one represents a general case but drops out the minor modification, while the latter one represents some special cases like Chern-Simons gravity. 

$A_\nu$ and $A_\mu$ are given by
\begin{equation}
    \label{convert_Mpv}
    \begin{aligned}
        A_\nu &= \frac{1}{M_{\rm PV}} [\alpha_\nu (0) -(1+z) \alpha_\nu (z)],\\
        A_\mu &= \frac{1}{M_{\rm PV}} \int_{0}^{z} \frac{(1+z')\alpha_\mu(z')}{\sqrt{\Omega_M (1+z')^3 + \Omega_\Lambda}},
    \end{aligned}
\end{equation}
where $z$ is redshift of the GW source. One can also rewrite the waveform in plus and cross polarizations via $h_{+}=(h_{\mathrm{L}}+h_{\mathrm{R}})/\sqrt{2}, h_{\times}=(h_{\mathrm{L}}-h_{\mathrm{R}})/\sqrt{2} i$~\cite{MTW}
\begin{equation}
    \label{waveform_PV}
    \begin{aligned}
        h_{+}^{\rm PV}(f) &= h_{+}^{\rm GR}(f) - h_{\times }^{\rm GR}(f) (i \delta h - \delta \Psi), \\
        h_{\times}^{\rm PV}(f) &= h_{\times}^{\rm GR}(f) + h_{+ }^{\rm GR}(f) (i \delta h - \delta \Psi).
    \end{aligned}
\end{equation}
This is the waveform we use in this work. For the background cosmological model, we adopt a flat Planck cosmology with parameters $ \Omega_{M}=0.308$, $\Omega_{\Lambda}=0.692$, $H_{0}=67.8 \mathrm{km} / \mathrm{s} / \mathrm{Mpc}$~~\cite{Planck1,Planck2}.

\section{Space-based GW Detectors \label{Sec3}}

\subsection{Basic Information: Configuration and Noise}
In this section we introduce the configurations and noise curves of three proposed space-based GW detectors, LISA, Taiji and TianQin. These are decisive factors for a detector's response to a coming GW signal. 

All the three detectors consist of a triangle of three spacecrafts, but they have different arm length, i.e., the separation between two spacecrafts. Arm length determines the sensitive frequency of a GW detector. Longer arm length corresponds to a lower frequency band (longer wavelength). LISA has an arm length of $2.5 \times 10^{6} ~\mathrm{km}$ and the designed sensitive frequency is from $10^{-4}$ to $1$ Hz~\cite{Amaro-Seoane2017}. Taiji's arm length is $3 \times 10^{6} ~\mathrm{km}$, which means Taiji is more sensitive to the lower frequency gravitational waves~\cite{Ruan2019}. TianQin's arm length is $1.7 \times 10^{5} ~\mathrm{km}$~\cite{Hu2018}, so it will be more sensitive at relative higher frequencies. This is consistent with the noise power spectral densities (PSDs) of these detectors. For LISA, we follow the new LISA design~\cite{Belgacem2019}, in which the PSD is given by
\begin{equation}
    S_{n}(f)=\frac{4 S_{\mathrm{acc}}(f)+S_{\mathrm{other}}} L\left[1+\left(\frac{f}{1.29 f_{*}}\right)^{2}\right], 
\end{equation} 
where $f_{*} = c/2\pi L$ is the transfer frequency of detector and $L$ is the arm length. The motion of LISA causes acceleration noise, which takes the form
\begin{widetext}
    \begin{equation}
        S_{\mathrm{acc}}(f)=\frac{9 \times 10^{-30} \mathrm{m}^{2} \mathrm{Hz}^{3}}{(2 \pi f)^{4}}\left[1+\left(\frac{6 \times 10^{-4} \mathrm{Hz}}{f}\right)^{2}\left(1+\left(\frac{2.22 \times 10^{-5} \mathrm{Hz}}{f}\right)^{8}\right)\right],
    \end{equation}
\end{widetext}
and other noise is 
\begin{equation}
    S_{\text {other }}=8.899 \times 10^{-23} \mathrm{m}^{2} \mathrm{Hz}^{-1}.
\end{equation}

For Taiji and TianQin, we employ a general noise curve for space-based GW detectors ~\cite{Huang2020,Liu2020}

\begin{equation}
    \begin{aligned}
        S_{n}(f)=&\left[\frac{S_{x}} L^{2}+\frac{4 S_{a}}{(2 \pi f)^{4} L^2} \left(1+\frac{10^{-4} \mathrm{Hz}}{f}\right)\right] \\
        &\times \left[1+\left(\frac{f}{1.29 f_{*}}\right)^{2}\right], 
    \end{aligned}
\end{equation}
where $\sqrt{S_a} = 3 \times 10^{-15} \mathrm{ms}^{-2} / \mathrm{Hz}^{1 / 2}$, $\sqrt{S_x} = 8 \times 10^{-12} \mathrm{m} / \mathrm{Hz}^{1 / 2}$ for Taiji, and $\sqrt{S_a} = 10^{-15} \mathrm{ms}^{-2} / \mathrm{Hz}^{1 / 2}$, $\sqrt{S_x} = 10^{-12} \mathrm{m} / \mathrm{Hz}^{1 / 2}$ for TianQin. 

\begin{figure}
    \includegraphics[width=0.5\textwidth]{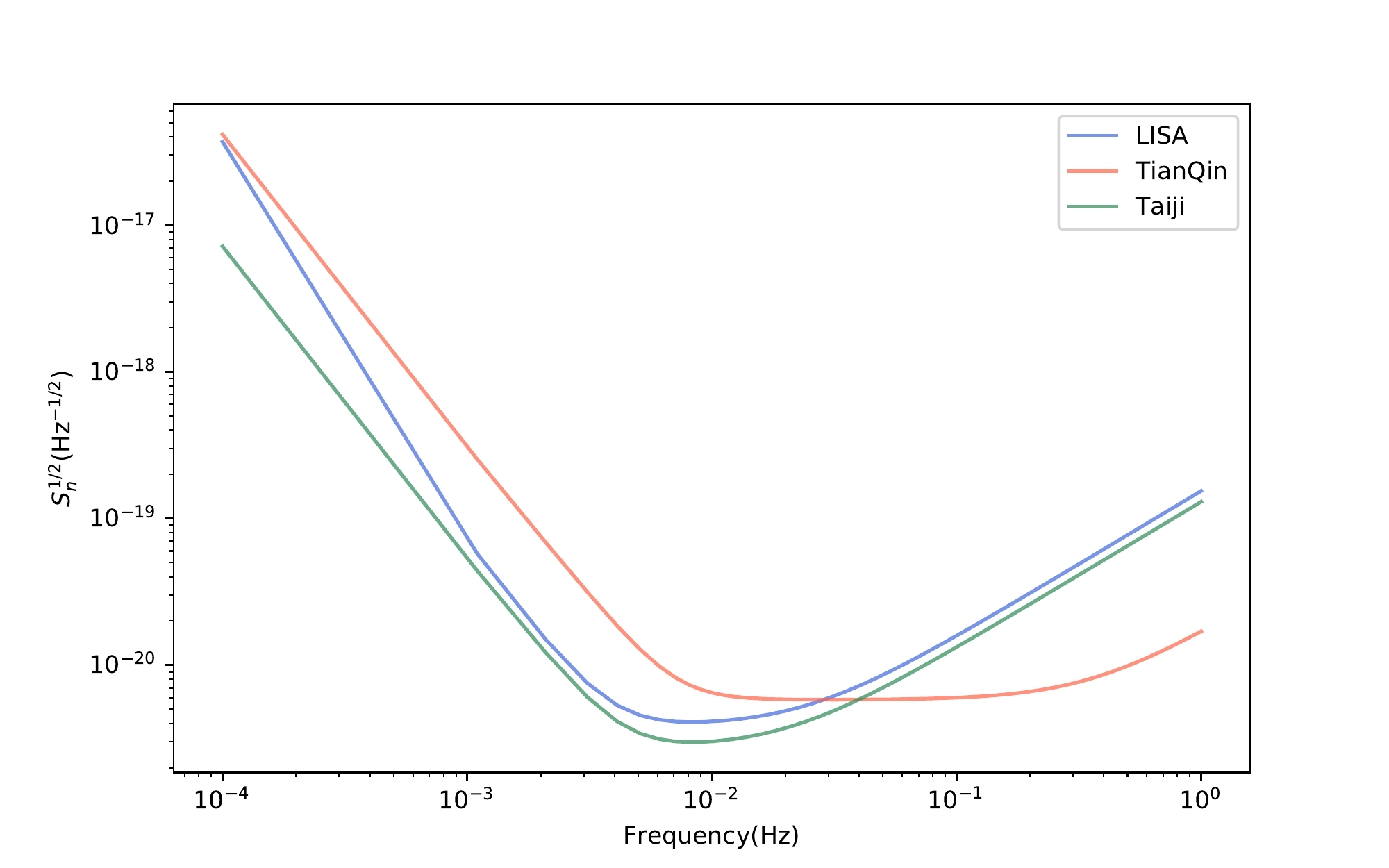}
    \centering
    \caption{\label{psd_plot} Noise power spectra of three space-based GW detectors. Blue, green, red lines represent LISA, Taiji and TianQin, respectively. Taiji and LISA have smaller noise in low frequency band, while TianQin is more sensitive to the relatively higher frequencies.}
\end{figure}

Their noise spectra is shown in Fig.~\ref{psd_plot}. As discussed before, LISA and Taiji are more sensitive than TianQin at lower frequency because of their longer arms, but less sensitive at higher frequency. 

\begin{figure}
    \subfigure[LISA and Taiji]{\includegraphics[width=0.4\textwidth]{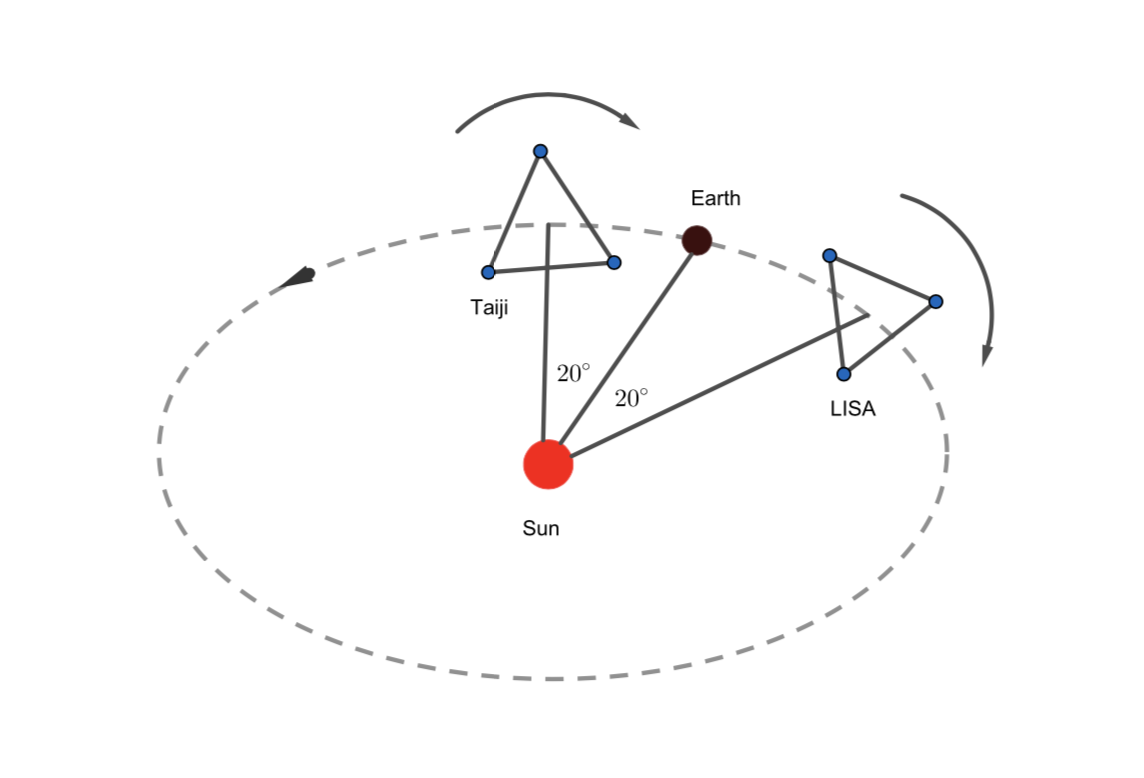}}
    \subfigure[TianQin]{\includegraphics[width=0.4\textwidth]{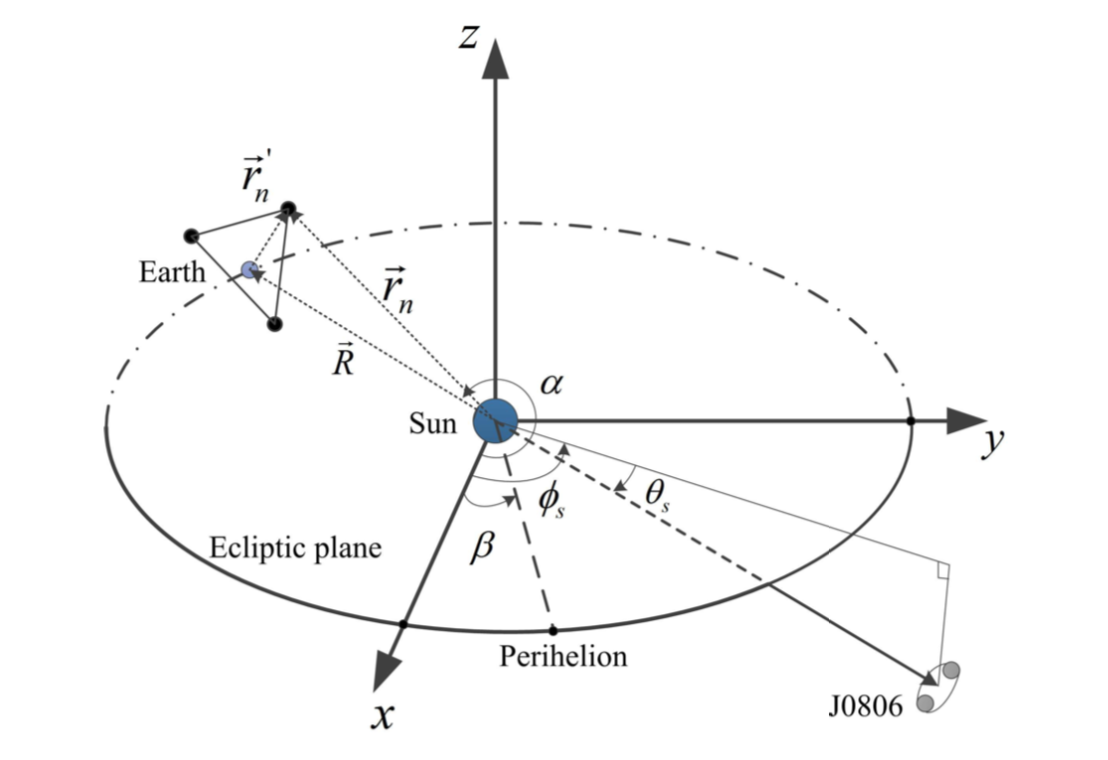}}
    \centering
    \caption{\label{orbits} Configuration of space-based GW detectors. Upper panel is the LISA-Taiji network configuration~\cite{Ruan2019}, in which LISA and Taiji take heliocentric orbits and separated by an angle of $40^{\circ}$. Lower panel is TianQin's orbit configuration in the heliocentric-ecliptic coordinate system~\cite{Hu2018}. The ecliptic plane is spanned by $x$ and $y$ axes. $x$ axis points toward the direction of the vernal equinox. $\beta$ is the longitude of the perihelion. Normal of TianQin's detector plane points to the reference source RX J0806.3+1527 whose coordinates in ecliptic frame is $(\theta_s, \phi_s)$. }
\end{figure}

In addition to arm length, three GW detectors also have different orbit designs. For instance, LISA's center of mass orbits around the Sun in ecliptic plane and the spacecrafts orbits their center of mass. Both of the two circular motions have the period of one year. Three spacecrafts constitute the shape of an equilateral triangle and the plane of the detector is tilted by $60^{\circ}$ with respect to the ecliptic~\cite{Amaro-Seoane2017}. The constellation falls behind the Earth by an angle of $\sim 20^{\circ}$. Taiji has a similar orbit, but it is ahead of the Earth by $20^{\circ}$. As shown in Fig.~\ref{orbits}, LISA and Taiji are far apart (about 0.7AU), by which GW localization could be improved~\cite{Ruan2019}. 
Considering circular orbits, the unit vectors along three arms in ecliptic frame can be derived. We define the $x$-$y$ plane as the ecliptic plane and $z$-axis as perpendicular to $x$-$y$ plane. Denoting $\mathbf{u_n}~(n=1,2,3)$ as the $n$-th arm defined in Fig.~2 of Ref.~\cite{Cutler1998}, it takes the form
\begin{equation}
    \label{ltvec}
    \begin{aligned}
        \mathbf{u_n} = (~&\frac{1}{2}\sin\alpha_n(t)\cos\phi(t)- \cos\alpha_n(t)\sin\phi(t),\\
        &\frac{1}{2}\sin\alpha_n(t)\sin\phi(t) + \cos\alpha_n(t)\cos\phi(t) ,\\
        &\frac{\sqrt{3}}{2} \sin\alpha_n(t)~),
    \end{aligned}
\end{equation}
with
\begin{equation}
    \alpha_{n}(t)=2 \pi t / T-\pi / 12-(n-1) \pi / 3+\alpha_{0},
\end{equation}
\begin{equation}
    \phi(t) = \phi_0 + 2\pi t/T,
\end{equation}
where $\alpha_0$ is a constant specifying the orientation of the arms at $t = 0$, $\phi_0$ specifies the detector's location at $t = 0$, and $T$ equals to one year. These vectors will be used in next subsection to calculate the instrument response.

As for TianQin (which is also shown in Fig.~\ref{orbits}), the orbit is more complex. Three spacecrafts orbit around the Earth, and the normal of the detector plane points to the reference source RX J0806.3+1527~\cite{Hu2018}. Previous works have derived the trajectory of TianQin in ecliptic frame: the $n$-th spacecraft's position vector $\mathbf{r_n(t)} = (x_n(t),y_n(t),z_n(t)),~n=1,2,3$ (shown in Appendix \ref{App1}). Arm direction vectors can be derived from $r_n(t)$. Considering $\mathbf{u_1}$ as an example, it is defined by
\begin{equation}
    \label{TianQinvec}
    \mathbf{u_1} = \frac{\mathbf{r_2} - \mathbf{r_1}}{|\mathbf{r_2}-\mathbf{r_1}|}.
\end{equation}
Thus, giving initial location and direction, the detectors' coordinates and arm direction vectors in ecliptic frame are determined. 

\subsection{Response}
In this section, we calculate space-based detectors' response to GWs. All azimuthal variables are defined in ecliptic frame.

Generally speaking, a GW detector's response $s(t)$ is a linear combination of GW's polarizations~\cite{GW1}
\begin{equation}
    \label{fhfh}
    s(t) = F_{+}h_{+}(t) + F_{\times}h_{\times}(t),
\end{equation}
where $h_{+}(t)$ and $h_{\times}(t)$ are plus and cross polarizations of GW. $F_{+}$ and $F_{\times}$ are antenna response functions, which are equal to the contraction of detector tensor $D^{ij}$ and GW polarization tensor $e^{\mathrm{A}}_{ij}$ with $\mathrm{A}=\{ +, \times \}$, i.e., 
\begin{equation}
    F_{\mathrm{A}} = D^{ij} e^{\mathrm{A}}_{ij},
\end{equation}
where $e^{\mathrm{A}}_{ij}$ in ecliptic frame is defined by a set of unit vectors $\{\hat{m}, \hat{n}, \hat{w}\}$ ~\cite{Cornish_2001,Liang2019},
\begin{equation}
    e_{i j}^{+}=\hat{m}_{i} \hat{m}_{j}-\hat{n}_{i} \hat{n}_{j}, \quad e_{i j}^{\times}=\hat{m}_{i} \hat{n}_{j}+\hat{n}_{i} \hat{m}_{j},
\end{equation}
with 
\begin{equation}
    \begin{aligned}
    \hat{m}=& (\cos \theta_{e} \cos \phi_{e} \cos \psi_{e}+\sin \phi_{e} \sin \psi_{e}\\
    & \cos \theta_{e} \sin \phi_{e} \cos \psi_{e}-\cos \phi_{e} \sin \psi_{e},\\
    & -\sin \theta_{e} \cos \psi_{e}),
    \end{aligned}
\end{equation}
\begin{equation}
    \begin{aligned}
    \hat{n}=&(-\cos \theta_{e} \cos \phi_{e} \sin \psi_{e}+\sin \phi_{e} \cos \psi_{e}\\
    & -\cos \theta_{e} \sin \phi_{e} \sin \psi_{e}-\cos \phi_{e} \cos \psi_{e},\\
    & \sin \theta_{e} \sin \psi_{e}),
    \end{aligned}
\end{equation}
\begin{equation}
    \hat{w}=\left(-\sin \theta_{e} \cos \phi_{e},-\sin \theta_{e} \sin \phi_{e},-\cos \theta_{e}\right),
\end{equation}
where $( \theta_e, \phi_e )$ are spherical coordinates in solar system with the ecliptic as $x$-$y$ plane and the Sun at center. $\psi_e$ is the polarization angle. $\hat{w}$ is propagation direction of GW, pointing from the source to the Sun.

\begin{figure}
    \includegraphics[width=0.5\textwidth]{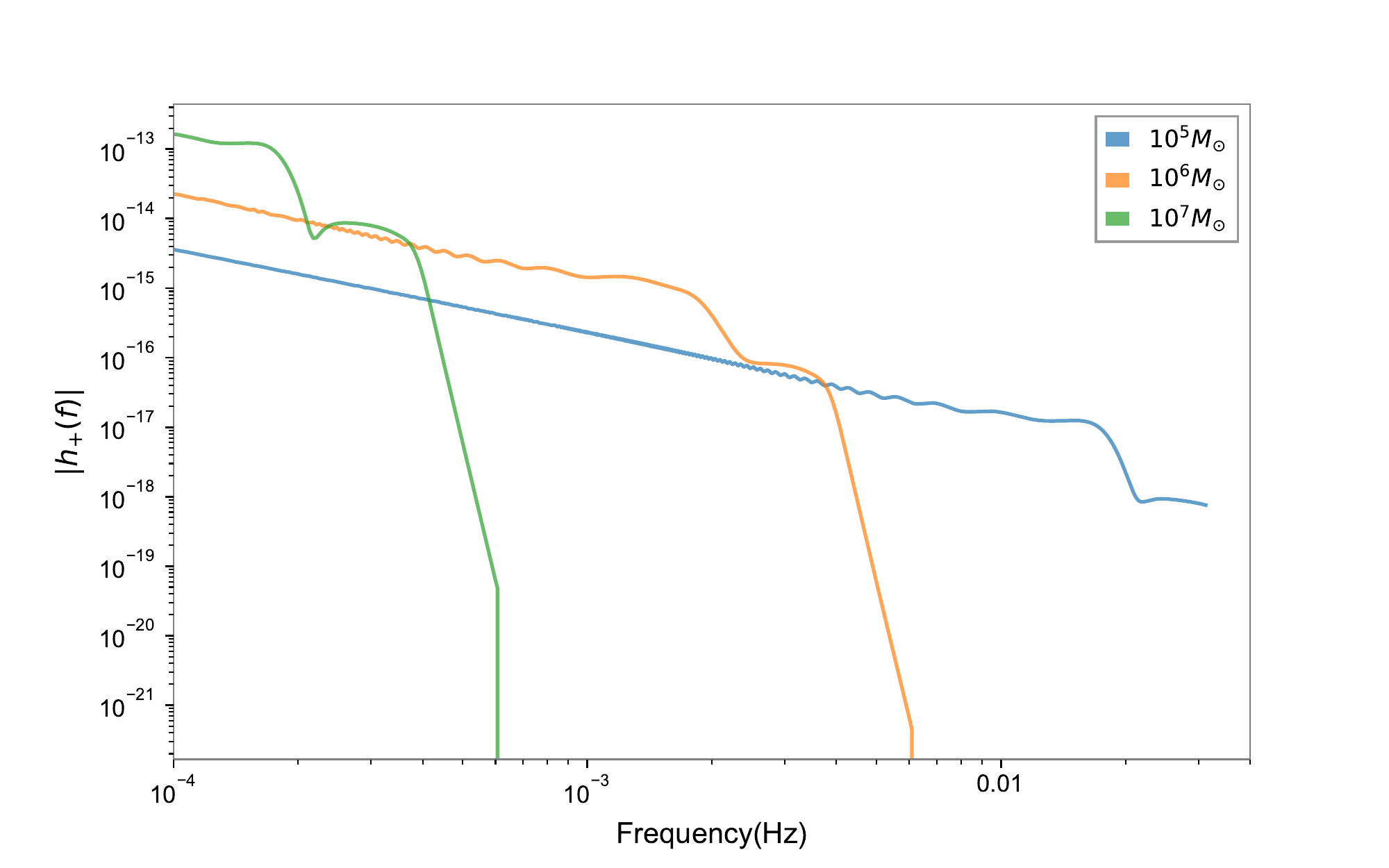}
    \centering
    \caption{\label{waveform} $|h_+ (f)|$ of GWs from different sources. Blue, orange and green lines are generated from SMBHBs with component masses of $5\times 10^{5} $,$5\times 10^{6} $,$5\times 10^{7} M_{\odot}$, respectively.  
    Here, we use the waveform template $\mathrm{IMRPhenomXHM}$.}
\end{figure}

The detector tensor, however, worths more discussions. Detector tensor is related to the tensor product of arm direction vectors. For ground-based GW detectors aiming at short-duration gravitational wave transient, arm direction vectors can be regarded as a constant during a GW event, thus detector tensor is also a constant. However, for space-based GW detectors whose objects are SMBHBs and EMRIs, observation often takes months to years. That is to say, detector tensor should be treated as a function of time, rather than a constant. In addition, since the wavelength of GWs is comparable to the physical arm length of detector (which is not satisfied for ground-based detectors), the GW frequency also makes a difference. In this case, we have~\cite{Cornish_2001,Liang2019} 
\begin{equation}
    D^{i j}(t;f)=\frac{1}{2}\left[\hat{u}^{i}(t) \hat{u}^{j}(t) T(f, \hat{u} \cdot \hat{w})-\hat{v}^{i}(t) \hat{v}^{j}(t) T(f, \hat{v} \cdot \hat{w})\right],
\end{equation}
where $\hat{u}^{i}(t)$ and $\hat{v}^{i}(t)$ are unit vectors along the arms of detector given in Eq.~(\ref{ltvec}) or (\ref{TianQinvec}). $T(f, \hat{u} \cdot \hat{w})$ is transfer function defined as
\begin{widetext}
\begin{equation}
    \begin{aligned}
    T(f, \hat{u} \cdot \hat{w})=\frac{1}{2} \left\{\operatorname{sinc}\left[\frac{f}{2 f^{*}}(1-\hat{u} \cdot \hat{w})\right] \exp \left[-i \frac{f}{2 f^{*}}(3+\hat{u} \cdot \hat{w})\right]\right.
    \left.+\operatorname{sinc}\left[\frac{f}{2 f^{*}}(1+\hat{u} \cdot \hat{w})\right] \exp \left[-i \frac{f}{2 f^{*}}(1+\hat{u} \cdot \hat{w})\right]\right\},
    \end{aligned}
\end{equation}
\end{widetext}
where $\mathrm{sinc}(x) \equiv{\sin x}/{x}$. Note that in low-frequency cases ($f \ll f_*$), transfer function tends to 1. The low-frequency approximation is widely used in previous works on LISA and we will also adopt this approximation. This is reasonable as the frequency of coalescence of SMBHBs is up-to $\sim 10^{-3}$Hz, while $f_*$ of the LISA, Taiji and TianQin detectors are $0.016,0.019$ and $0.28$Hz, respectively. We plot GW waveform from SMBHBs of different masses in frequency domain in Fig.~\ref{waveform}, from which we find the low-frequency approximation works well for SMBHBs with masses higher than $10^6 M_{\odot}$. In this work, we employ a higher cut-off frequency $10^{-2}$Hz, above which data is not included in analysis.

Because of the three-arm design, a single space-based GW detector can output two independent strains~\cite{Cutler1998}. Thus, a detector corresponds to two detector tensors. In accordance with time delay interferometry, one can define two detector tensors $D_a^{i j}, D_e^{i j}$ as~\cite{Marsat2020}
\begin{equation}
    \begin{aligned}
        D_a^{i j} &= \frac{1}{6}(u_1^i u_1^j - 2u_2^i u_2^j +u_3^i u_3^j), \\
        D_e^{i j} &= \frac{\sqrt{3}}{6}(u_1^i u_1^j -u_3^i u_3^j ),
    \end{aligned}
\end{equation}
where $u_1^i, u_2^i,u_3^i$ are arm direction vectors for three arms. Note that, this formula is written in the low-frequency limit.

When performing Bayesian analysis, we need GW data in frequency domain. It is difficult to do Fourier transformation directly to Eq.~(\ref{fhfh}), due to antenna pattern functions' dependency on time. To solve this problem, we adopt stationary phase approximation (SPA). In SPA, frequency domain response can be written as 
\begin{equation}
    \label{SPAresponse}
    \tilde{s}(f) = F_{+}\left[ t(f)\right] \tilde{h}_{+}(f) + F_{\times}\left[ t(f)\right]\tilde{h}_{\times}(f),
\end{equation}
that is to say, we can change $F_{+}(t)$ into $F_{+}\left[t(f)\right]$ as a replacement of Fourier transform. The expression of $t(f)$ is given in Appendix \ref{App2}. Here, a \textit{tilde} denotes the quantity in frequency domain.

Note that waveform in frequency domain should include the time delay to the Sun by adding an extra phase term as follows,
\begin{equation}
    \label{AddPhase}
    \tilde{h}_{+,\times}(f)  = \mathcal{F}\left[h_{+,\times}(t)\right] \exp{\left[-2\pi i f (\frac{\hat{w}\cdot \vec{r}}{c} + t_c -t_0 )\right]},
\end{equation}
where $\mathcal{F}$ means Fourier transform, $t_c$ is coalescence time and $t_0$ is the start time of data. 

\section{Methodology \label{Sec4}}
\subsection{Bayesian Method}
Bayesian method is one of the most widely-used ways of parameter estimation in GW astronomy~\cite{Thrane2019}. Given observed data and prior distributions of parameters, one can obtain the posterior distribution by 
\begin{equation}
    \label{BayesianTheory}
    p(\vec{\vartheta} | \vec{d}(t))=\frac{p(\vec{d}(t) | \vec{\vartheta}) p(\vec{\vartheta} )}{p(\vec{d}(t))},
\end{equation}
where $\vec{d}(t)$ is observed data and $\vec{\vartheta}$ is parameter set. The denominator, evidence, is often ignored since it is a normalization constant if we only care about the distribution of parameters. We define inner product between two strains as
\begin{equation}
    \left\langle\tilde{a}(f) | \tilde{b}(f)\right\rangle=4 \Re \int_{0}^{\infty} \frac{\tilde{a}(f) \tilde{b}^{*}(f)}{S_{n}(f)} \mathrm{d} f,
\end{equation}
where a \textit{star} denotes complex conjugate. $S_{n}(f)$ is the PSD of the detector.  The likelihood, $p(\vec{d}(t) | \vec{\vartheta}, H)$,  takes the form 
\begin{equation}
    p(\vec{d}(t) | \vec{\vartheta}, H)=\exp \left[-\frac{1}{2} \sum_{i=1}^{N}\left\langle\tilde{n}_{i}(f) | \tilde{n}_{i}(f)\right\rangle\right],
\end{equation}
on the assumption that the noise is Gaussian~\cite{Thrane2019}. Here, the subscript $i$ denotes the $i$-th data strain and $\tilde{n}_{i}(f)$ is the noise. For the $i$-th strain that contains data $\tilde{d}_{i}(f)$, we simply have 
\begin{equation}
    \tilde{d}_{i}(f) = \tilde{s}_{i}(f) + \tilde{n}_{i}(f),
\end{equation}
where $\tilde{s}_{i}(f)$ is detector's response to GW signals. Thus, the likelihood can be written as 
\begin{widetext}
\begin{equation}
    p(\vec{d}(t) | \vec{\vartheta}, H)= \exp \left[-\frac{1}{2} \sum_{i=1}^{N}\left\langle\tilde{d}_{i}(f)-\tilde{s}_{i}(f, \vec{\vartheta}) | \tilde{d}_{i}(f)-\tilde{s}_{i}(f, \vec{\vartheta})\right\rangle\right].
\end{equation}
\end{widetext}
If the prior probability densities are also set, we can obtain the posterior distribution of parameters theoretically. Some numerical ways are developed to generate the posterior samples for given data and likelihood, including Markov-chain Monte Carlo method and Nested sampling method~\cite{Thrane2019}. In this work, we employ a multimodal nested sampling algorithm Multinest~\cite{Multinest,NestedSampling}. Nested sampling works with a set of live points generated from prior distributions. After each iteration, the point with the lowest likelihood will be abandoned and the new samples with higher likelihood will be generated. In the end, those live points will be mapped to posterior samples. 

Several tools for Bayesian parameter estimation in GW astronomy have been developed~\cite{LALInference,Biwer2019,Ashton2019}. We adopt and modify the Python toolkit Bilby~\cite{Ashton2019} in this work with sampler PyMultiNest~\cite{PyMultinest}. Codes for this paper could be found in \href{https://github.com/Li-mz/bilby/tree/SpaceInterferometer}{our Github repository}.

\subsection{Waveform and Parameters}
In this section, we clarify the parameters and the GW waveform used in this work.

As mentioned in Eq.~(\ref{waveform_PV}), GW waveform in PV gravity is GR waveform with phase and amplitude modifications. Thus, what we need to do is to choose an appropriate GR waveform template. Previous studies have shown that the public IMRPhenom waveform with high harmonics works fairly in Bayesian analysis~\cite{London2018}. Subsequent works emphasize that the high harmonics play an important role in parameter estimation for space-based GW detectors~\cite{Marsat2020,Baibhav_2020}. For these reasons, we choose IMRPhenomXHM~\cite{IMRXHM2020}, a frequency domain model for the GW of non-precessing black-hole binaries with high harmonics available. One can decompose waveform into spherical harmonic modes~\cite{Blanchet2014}
\begin{equation}
    \begin{aligned}
    h_{+} &=\sum_{\ell, m}h_{\ell m, +} =\frac{1}{2} \sum_{\ell, m}\left(_2Y_{\ell m} h_{\ell m}+_{-2} Y_{\ell m}^{*} h_{\ell m}^{*}\right) ,\\
    h_{\times} &=\sum_{\ell, m}h_{\ell m, \times} = \frac{i}{2} \sum_{\ell, m}\left(_{2} Y_{\ell m} h_{\ell m}-_{-2} Y_{\ell m}^{*} h_{\ell m}^{*}\right),
    \end{aligned}
\end{equation}
where $_2 Y_{\ell m}$ is spin-weighted spherical harmonics~\cite{Blanchet2014}. Except for the dominant term $(\ell, m) = (2,2)$, we also adopt higher modes including $(\ell, m) = (2,1), (3,3), (4,4), (5,5)$ in our analysis. Note that different modes correspond to different frequency components of GW, thus the function $t(f)$ from SPA differs from modes to modes. We have 
\begin{equation}
    t_{\ell m}(f) = t_{22}(2f/m)
\end{equation}
where $t_{22}(f)$ is given in Appendix \ref{App2} and Eq.~(\ref{SPAresponse}) should be rewritten as 
\begin{equation}
    \tilde{s}(f) = \sum_{\ell, m} F_{+}\left[t_{\ell m}(f)\right]\tilde{h}_{\ell m,+}(f) + F_{\times}\left[t_{\ell m}(f)\right]\tilde{h}_{\ell m,\times}(f).
\end{equation}

In general, GWs from compact binary black holes have fifteen basic parameters: masses of two black holes, spins of two black holes (six components in total), luminosity distance $d_L$, coalescence time $t_c$, coalescence phase $\phi$, inclination angle $\iota$, polarization angle $\psi_e$, and source direction which in our work is ($\phi_e, \theta_e$). There are other two parameters in parity-violating gravity that specify velocity and amplitude birefringence respectively. As discussed in Sec.~\ref{Sec2}, to investigate the constraint on parity asymmetry, we consider two cases. (1) GW waveform with only velocity birefringence. We ignore amplitude modification, since it is a minor factor compared with phase modification. (2) GW waveform with only amplitude birefringence, as some gravity theories predict only amplitude birefringence.

In PV gravity, $\delta h$ and $\delta \Psi$ are the two modification terms. We choose $A_\mu /H_0$ and $-A_\nu$ as additional parameters in waveform, and denote them as $A$ and $B$, respectively. The phase and amplitude modifications can be written as
\begin{equation}
    \begin{aligned}
        \delta h(f) &= B (\pi f), \\
        \delta \Psi(f) &= A (\pi f)^2.
    \end{aligned}
\end{equation}
The posterior distributions of $A$ and $B$ can be easily converted to $M_{\mathrm{PV}}$ through Eq.~(\ref{convert_Mpv}).

A 16-dimensional full Bayesian analysis is extremely computational expensive, especially when higher modes are taken into consideration and several data strains are included (note that one detector produces two data strains). To lessen computation burden, we only consider zero-spin black holes, which means we have 9 parameters in GR and 1 additional modification parameter for PV gravity. The major effects of velocity and amplitude birefringence take place during propagation, so ignoring spins will not produce significant influence on our conclusions of constraints on PV gravity. Plus, employing non-spinning GW templates has negligible impact on sky localization, as previous studies suggest~\cite{SpinParaEst,Singer2016}.

Prior distributions of the remaining parameters are given as follows:
\begin{list}{}{}
\item[$\diamond$] Component masses: uniform distribution between $10^5~M_{\odot}$ and $10^7~M_{\odot}$.
\item[$\diamond$] Luminosity distance: uniform distribution between $10^3~\mathrm{Mpc}$ and $10^5~\mathrm{Mpc}$.
\item[$\diamond$] Coalescence time: uniform distribution between $t_c - 10~\mathrm{s}$ and $t_c + 10~\mathrm{s}$, where $t_c$ is the coalescence time of our injection.
\item[$\diamond$] Coalescence phase: uniform distribution in $[0, 2\pi]$.
\item[$\diamond$] Polarization angle: uniform distribution in $[0, 2\pi]$.
\item[$\diamond$] Inclination angle: sine distribution in $[0, \pi]$.
\item[$\diamond$] Source direction: uniform distribution in the sky, i.e., uniform distribution for $\phi_e$ and $\cos \theta_e$.
\item[$\diamond$] A: uniform distribution in $[-10^3~\mathrm{Hz}^{-2}, 10^3~\mathrm{Hz}^{-2}]$.
\item[$\diamond$] B: uniform distribution in $[-10^2~\mathrm{Hz}^{-1}, 10^2~\mathrm{Hz}^{-1}]$.
\end{list}

\section{Localization ability of detector networks \label{Sec5}}
In this section, we show the results of GW source localization given by different detector networks. We consider three cases: LISA, LISA-Taiji network and LISA-TianQin network. We first show parameters can be correctly estimated with the Bayesian framework, then present GW localization of sources in different direction.

\begin{figure*}
    \centering
    \includegraphics[width=1\textwidth]{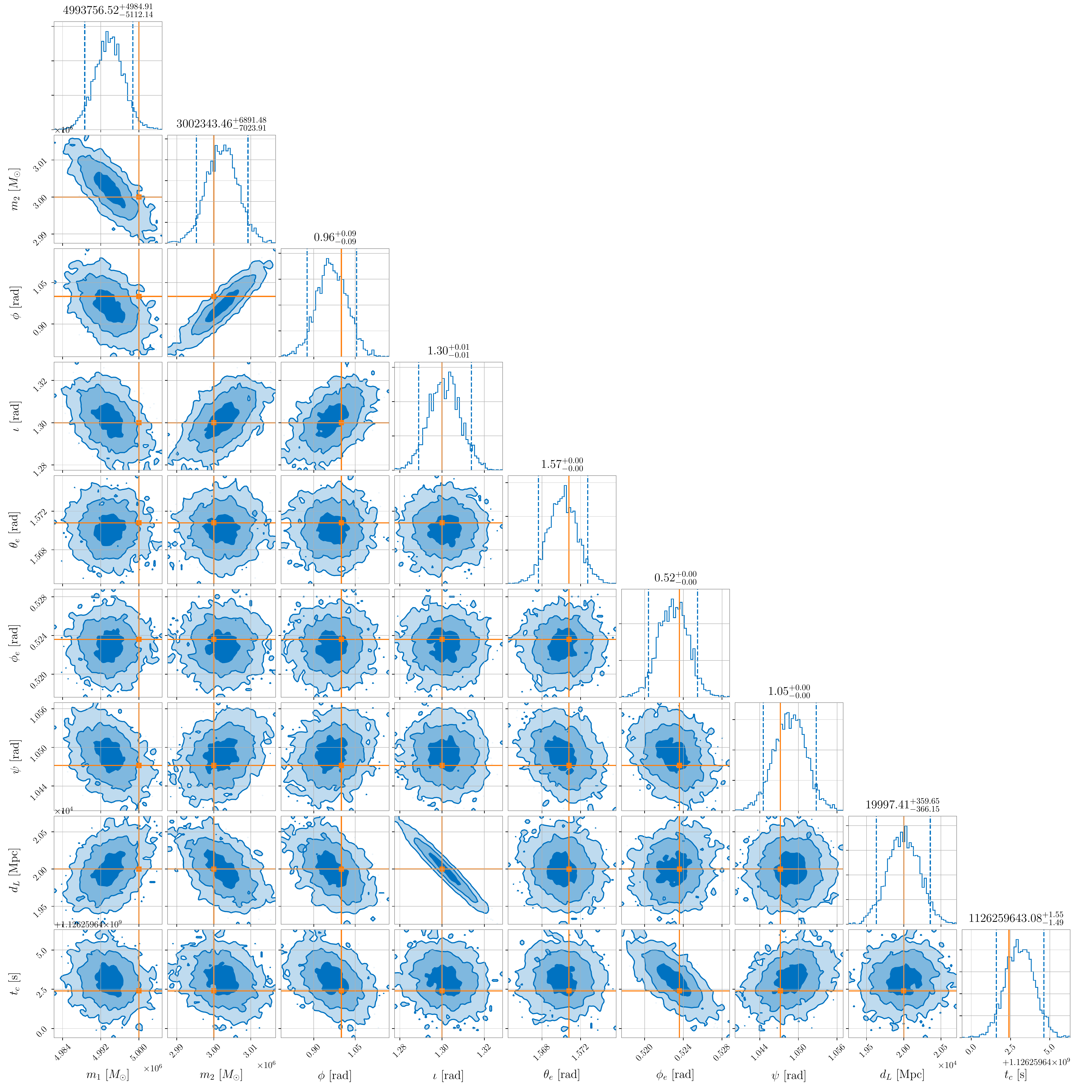}
    \caption{\label{LISA_GR_theta90_corner} Posterior distributions generated with single LISA observation. The yellow solid lines are injected values, and the blue dashed lines are 5\% and 95\% percentiles.
    }
\end{figure*}

\begin{figure*}
    \centering
    \includegraphics[width=1\textwidth]{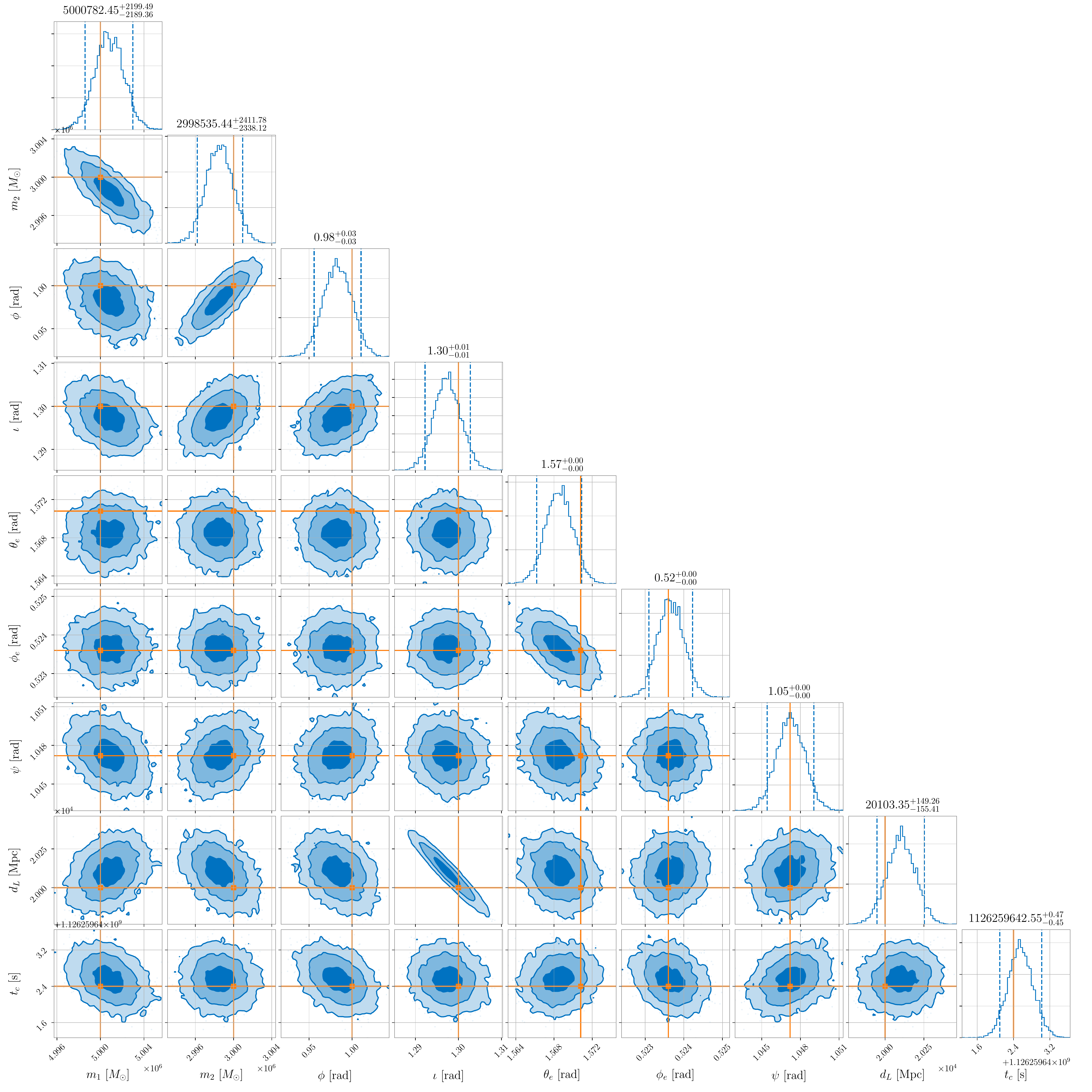}
    \caption{\label{LISA_Taiji_GR_theta90_corner} Posterior distributions generated with LISA and Taiji joint observation. The yellow lines are injected values, and blue dashed lines are 5\% and 95\% percentiles. 
    }
\end{figure*}

We simulate $2^{18}$ seconds (about 3 days) long GW data of an SMBHB with masses at order of $10^6 M_\odot$ and luminosity distance of $20$Gpc. Sampling frequency is set to $1/16$ Hz, which corresponds to the nyquist frequency of $0.03125$ Hz. This is consistent with the $0.01$ Hz cut-off. In order to cross-check the stability of the results, we have also considered the cases with sampling frequencies of $1/8$ Hz and $1/64$ Hz, and found the consistent results. With parallel computing using 16 processes, it takes the sampler 10 hours to generate the posterior samples for one detector, and 24 hours for joint observation of two detectors. As an illustration, we show the corner plots of LISA and LISA+Taiji network in Fig.~\ref{LISA_GR_theta90_corner} and \ref{LISA_Taiji_GR_theta90_corner}. The signal-to-noise ratio in LISA is higher than 500, which enables injected parameters to be correctly reconstructed. Some common correlations between parameters are also shown, e.g., component masses $m_1$ and $m_2$, luminosity distance $d_L$ and inclination $\iota$, component masses and phase $\phi$. Note that, the error bars of joint observation are reduced compared with a single detector, which implies that joint observation could significantly improve the parameter constraints. 

GW source localization depends on the time difference of GW signal's arrival in each detector, which is called triangulation information. However, sources in some specific directions produce much weaker triangulation information, which makes localization difficult. For example, overhead binaries~\cite{Baibhav_2020}, from $\theta_e = 60^{\circ}$ and $\phi_e$ close to LISA's mass center's $\phi_e$. The distances from the source to the three spacecrafts of LISA are roughly equal because LISA's detector plane is tilted by $60^{\circ}$ with respect to the ecliptic plane. Therefore, a single LISA may fail to localize the source in such a direction if observation duration is not long enough. By contrast, different detectors in a detector network may be separated by at most 0.7AU and can avoid the overhead binaries problem. 

\begin{figure*}
    \subfigure[$\theta_e = 30^{\circ}$]{\includegraphics[width=0.5\textwidth]{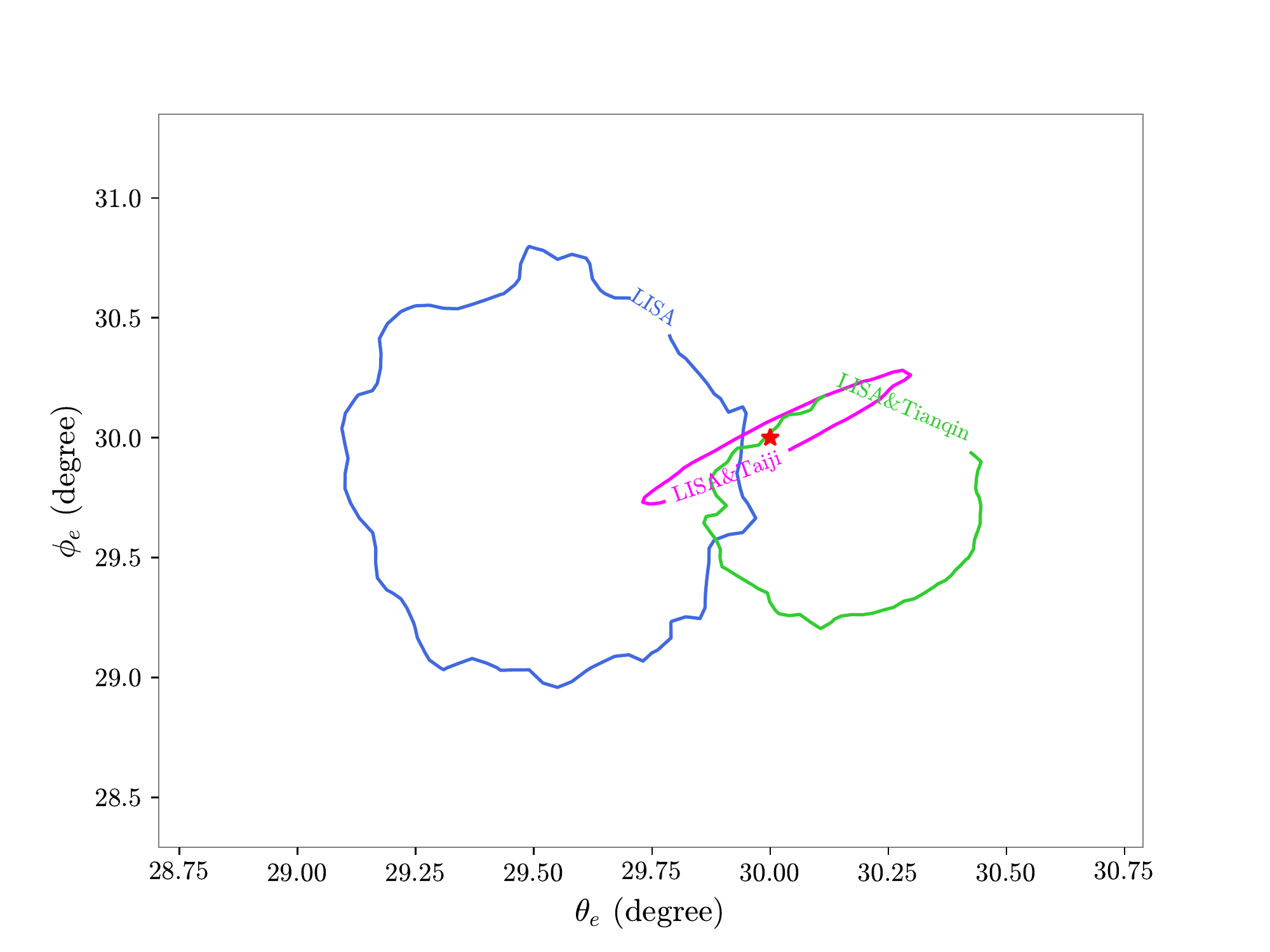}}
    \hspace{-.15in}
    \subfigure[$\theta_e = 90^{\circ}$]{\includegraphics[width=0.5\textwidth]{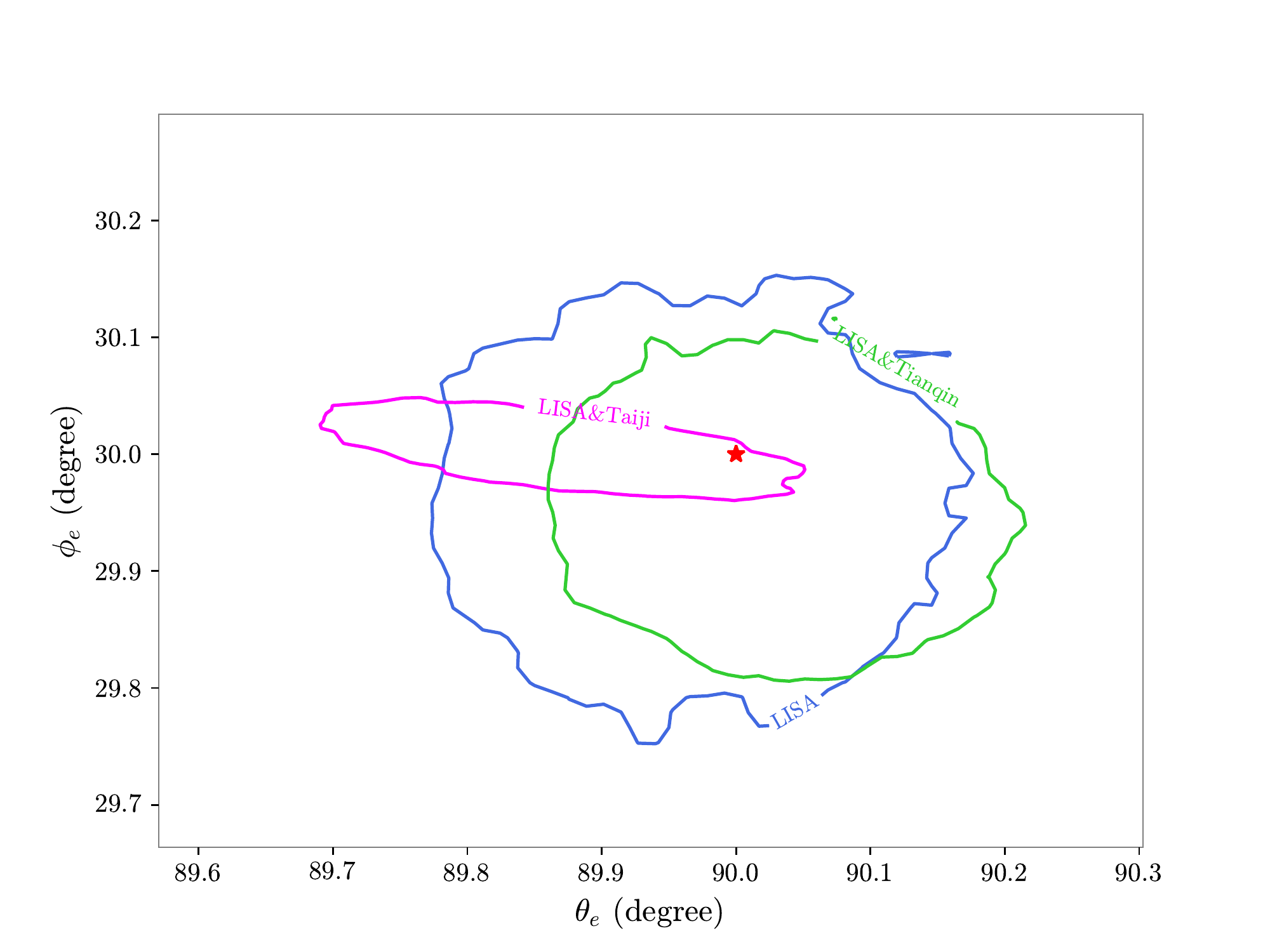}}
    \subfigure[$\theta_e = 60^{\circ}$]{\includegraphics[width=0.65\textwidth]{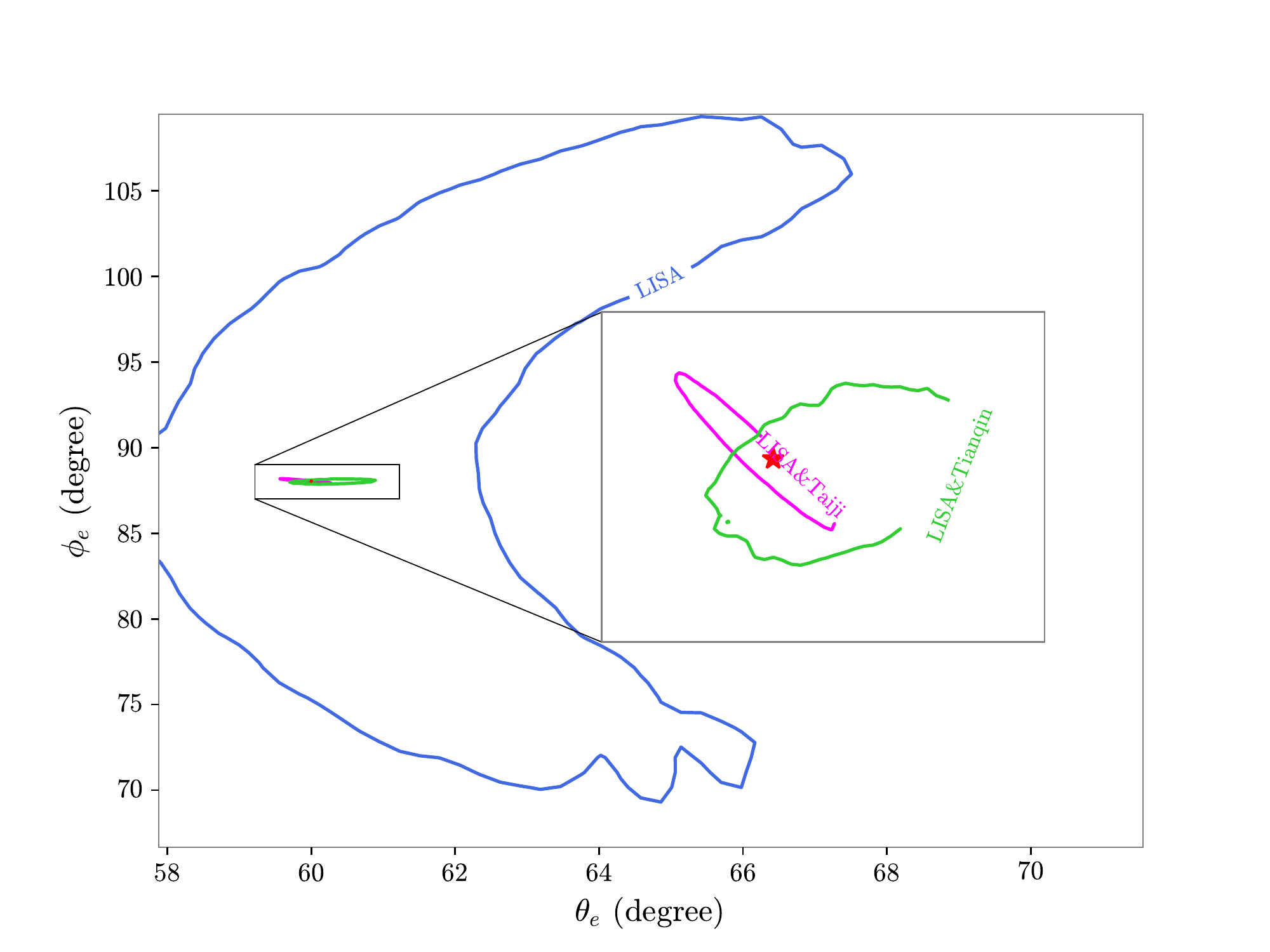}}
    \centering
    \caption{\label{localization_contour} The 90\% credible contours of posterior distribution of $(\theta_e, \phi_e)$ for GW sources in three directions. Results of different detector networks are plotted in different color. For each panel, the true location is indicated by a red star. 90\% credible areas for the case with $\theta_e = 30^{\circ}$ are $0.54~\mathrm{deg}^2$, $0.034~\mathrm{deg}^2$, $0.20~\mathrm{deg}^2$ for LISA, LISA+Taiji network, LISA+TianQin network, respectively. For the case with $\theta_e = 60^{\circ}$, they are $160.3~\mathrm{deg}^2$, $0.035~\mathrm{deg}^2$ and $0.22~\mathrm{deg}^2$, respectively. For the case with $\theta_e = 90^{\circ}$, they are $0.11~\mathrm{deg}^2$, $0.019~\mathrm{deg}^2$ and $0.079~\mathrm{deg}^2$, respectively. }
\end{figure*}

We simulate three GW sources with $\theta_e = 30^{\circ}, 60^{\circ}, 90^{\circ}$. The 90\% credible areas of posterior distributions of $(\theta_e, \phi_e)$ are shown in Fig.~\ref{localization_contour}. As anticipated, due to the much longer baselines, the detector networks could significantly reduce the localization area. Typical 90\% credible area of a single LISA is $\mathcal{O}(10^{-1}) ~\mathrm{deg}^2$, while for detector networks it is $\mathcal{O}(10^{-2}) ~\mathrm{deg}^2$. In the special $\theta_e = 60^{\circ}$ case, detector networks can bring an improvement of four orders of magnitude. 

Note that, LISA+Taiji network gives stronger improvements than LISA+TianQin network, which is understandable. As mentioned above, in this article, we set a high frequency cut-off of $0.01$ Hz. From Fig.\ref{psd_plot}, we find that in this frequency range, Taiji has the much lower noise level, which can produce the larger SNRs in this frequency band for the given event, hence the improvement on source localization is more distinct. On the other hand, the main advantage of TianQin is at the higher frequency range of $f>0.03$Hz, which is more sensitive to detect the BBHs with component mass less than $10^{5}M_{\odot}$.

\section{Constraints on PV gravity \label{Sec6}}
We will show the constraints on PV gravity given by detector networks in this section. In Sec.~\ref{Sec4}, we defined two parameters $A$ and $B$ in parity-violating GW waveforms and explained two cases to consider. Here, we inject GW signals from SMBHBs with the same masses and distance as in previous section, and set $A=B=0$ in our fiducial model. In the Bayesian analysis, we add the PV parameters to the parameter set and obtain their distributions. Note that, the expected values of these PV parameters are zero -- our intention is to investigate the capabilities of constraining PV gravity of detector networks, so we focus on the error bars of PV parameters, which are not sensitive to the injected values. 

\begin{figure*}
    \subfigure[velocity birefringence]{\includegraphics[width=0.5\textwidth]{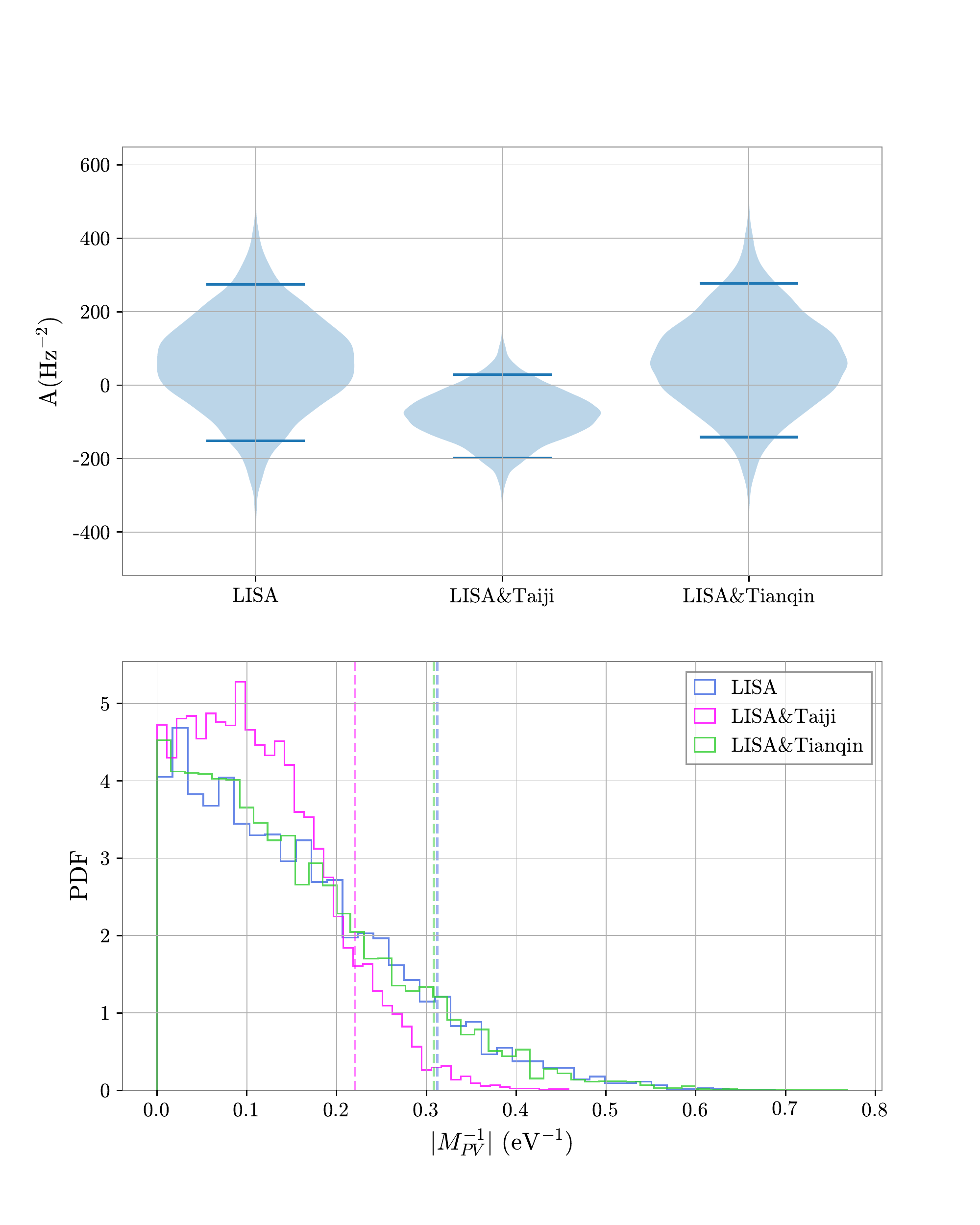}}
    \hspace{-.15in}
    \subfigure[amplitude birefringence]{\includegraphics[width=0.5\textwidth]{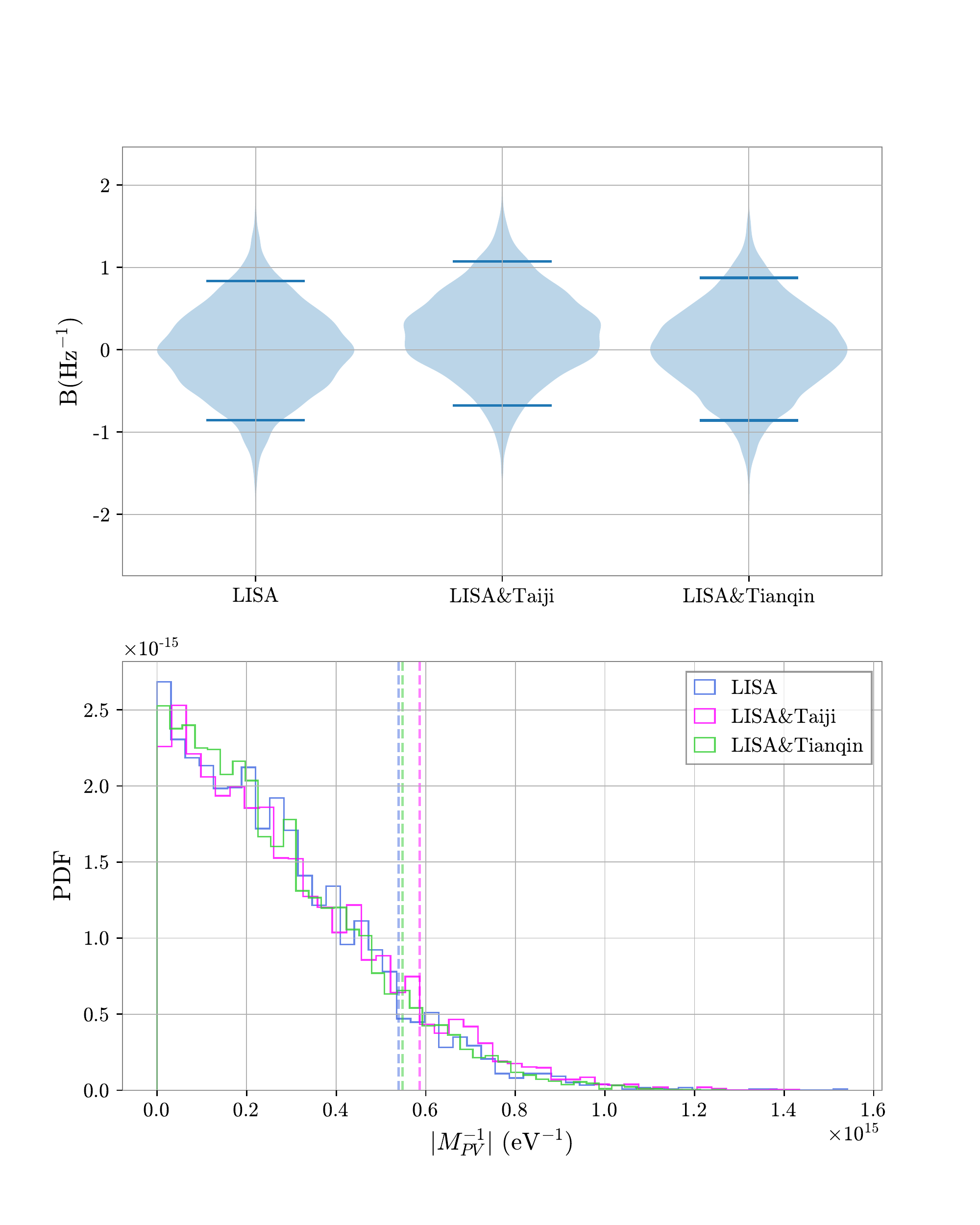}}
    \centering
    \caption{\label{PV_posterior} 
    Upper panels show the violin plots for effective PV parameters $A$ and $B$ and lower panels show the distribution of $M_{\mathrm{PV}}$ derived from different cases. Note that, 90\% percentiles are plotted in dashed lines. For velocity birefringence, 90\% percentiles correspond to $3.20$eV, $4.53$eV, $3.24$eV for LISA, LISA+Taiji network, LISA+TianQin network, respectively. While for amplitude birefringence, 90\% percentiles are $1.85\times 10^{-15}$eV, $1.70\times 10^{-15}$eV, $1.82\times 10^{-15}$eV, respectively.}
\end{figure*}

With an additional PV parameter, the sampling time increases by roughly 50\%. Upper panel of Fig.~\ref{PV_posterior} shows the violin plots of posterior distribution of effective PV parameters. Also, $M_{\mathrm{PV}}$ can be calculated by effective PV parameters via Eq.~(\ref{convert_Mpv}) and is showed in the lower panel. Note that injected PV parameters are zero and the theoretical $M_{\mathrm{PV}}$ should be infinite, hence we plot distribution of $M_{\mathrm{PV}}^{-1}$ instead. Taking the 90\% percentiles of $M_{\mathrm{PV}}^{-1}$ as lower limit of $M_{\mathrm{PV}}$ in 90\% credible level, velocity birefringence effect gives $\mathcal{O}(1)$~eV and amplitude birefringence effect gives $\mathcal{O}(10^{-15})$~eV. It is reasonable that velocity birefringence effect follows a higher constraint than amplitude birefringence because the its physical effect is much stronger.

Compared with the constraints given by ground-based GW detectors~\cite{Wang2020,Zhao2019,Alexander_2018,okounkova2021constraining}, limits given by space detectors are not strong. For example, using LIGO-Virgo detections, Ref.~\cite{Wang2020} gives $M_{\mathrm{PV}}>0.07$ GeV by constraining velocity birefringence, and Ref.~\cite{okounkova2021constraining} gives $M_{\mathrm{PV}}>10^{-13}$ eV by constraining amplitude birefringence. We summarize the known constraints on PV gravity in Table~\ref{tab1}. As indicated in Eq.~(\ref{proptof}), the amplitude and phase modification in PV gravity are proportional to the GW frequency and square of the frequency, respectively. Sensitive frequency of a ground-based GW detector can be 5-6 orders of magnitude larger than space-based GW detectors, so the weaker limits are reasonable. 

\begin{table}[]
    \resizebox{0.5\textwidth}{!}{%
    \begin{tabular}{|c|c|}
    \hline
    \textbf{Method}                                                                             & \textbf{Lower Limit of $\bm{M_{\mathrm{PV}}}$} \\ \hline
    \begin{tabular}[c]{@{}c@{}}LIGO-VIRGO detections \\ velocity birefringence\cite{Wang2020}\end{tabular} & $0.07$~GeV                   \\ \hline
    \begin{tabular}[c]{@{}c@{}}LIGO-VIRGO detections \\ amplitude birefringence\cite{okounkova2021constraining}\end{tabular} & $10^{-13}$~eV       \\ \hline
    GW speed\cite{GWspeed1,GWspeed2}                                                            & $10$~eV                       \\ \hline
    Solar system tests\cite{solarsystem}                                                        & $2\times 10^{-13}$~eV         \\ \hline
    Binary pulsar\cite{binarypulsar1,binarypulsar2}                                             & $5\times 10^{-10}$~eV       \\ \hline
    \end{tabular}%
    }
    \caption{Constraint on parity-violating energy scale from different tests. }
    \label{tab1}
\end{table}

Unlike source localization, there is no statistically significant improvement of constraining PV gravity if we use a detector network. Since detector networks provide much longer baselines and thus the triangulation information is enhanced, detector networks can greatly improve the localization capability. However, the information of parity violation lies in the arrival time or amplitude difference of left- and right- hand polarizations, which cannot be significantly changed by detector networks, in comparison with an individual detector. Therefore, joint observations do not bring a significant improvement.

\section{Conclusions and Discussions \label{Sec7}}

The gravitational-wave signals, produced by the coalescence of compact binaries, provide the excellent opportunities to study the abundant physical processes and test the fundamental properties of gravity in the strong gravitational fields. In addition to various ground-based GW detectors, several space-based detectors, including LISA, Taiji and TianQin, are expected to be launched in the near future. They are sensitive to the GW signals at lower frequency bands, and will open a new window for the GW astronomy. In particular, in comparison with the individual detectors, detector networks consisting of several detectors might significantly improve the constraints of various parameters.
In this article, by applying Bayesian analysis, we investigate the capabilities of space-based detector networks, and consider two cases as the examples. The first case is to localize the GW sources with detector networks, and the second is to constrain the parity symmetry of gravity with GWs.

As well known, source localization is an important aspect of GW astronomy as it helps to identify the host galaxy of the source and directs observations of electromagnetic emission. In this work, we investigate the possible improvement of GW source localization with the potential observations of future detector LISA, as well as detector networks consist of LISA, Taiji and TianQin projects. In analysis, we first simulate GW signals with the waveform template IMRPhenomXHM, and inject them into various detectors. Then, employing the modified Bilby package, we use Bayesian method to estimate physical parameters of the compact binaries and constrain the parameters of source position. We find that a detector network can improve the localization area by one order of magnitude in a three-day observation of compact binaries of $\mathcal{O}( 10^6)~\mathrm{M_{\odot}}$. For GW sources in some special directions, a detector network is crucial to the successful localization. 

In the second case of testing gravity with GWs, we extend our previous works on testing the parity symmetry of gravity with GWs produced by the stellar-mass compact binaries to the case with SMBHBs.
By the similar analysis, we constrain the parameters which quantify the velocity birefringence and amplitude birefringence effects in PV gravity. We find that the individual space-based GW detectors and the detector networks can give the similar constraints: i.e., the lower bound of the PV energy scale $M_{\rm PV}\gtrsim\mathcal{O}(1)$eV by constraining the velocity birefringence effect of GWs and $M_{\rm PV}\gtrsim\mathcal{O}(10^{-15})$eV by constraining the amplitude birefringence effect of GWs. Since the space-based detectors are sensitive to the GW signal of lower frequencies, this bound is weaker than that derived from the observations of ground-based GW detectors.

At the end of this paper, we should mention that we have to simplify the calculation in the following aspects due to the complexity of space-based GW detector's response and nested sampling's computational burden. First, we adopt only three-day GW signals for analysis, which is much less than the realistic duration of future GW detection. Second, in Bayesian analysis, we use non-spinning GW waveform to reduce the parameter dimensionality. Third, in order to transfer the responses of detectors from time domain to frequency domain, we adopt the SPA to simplify our calculation. We should emphasize, these are common problems in community when it comes to Bayesian analysis of GW signal of space-based detectors, which should be overcome by various techniques in future works.  

~

~

~

\begin{acknowledgments}
We would like to thank Yifan Wang and Yiming Hu for several helpful discussions. This work is supported by NSFC No.11773028, 11633001, 11653002, 11603020, 11903030, the Fundamental Research Funds for the Central Universities under Grant Nos: WK2030000036 and WK3440000004, the Strategic Priority Research Program of the Chinese Academy of Sciences Grant No. XDB23010200, and the China Manned Space Program through its Space Application System.
\end{acknowledgments}

~

~

\appendix

\section{TianQin's Orbit\label{App1}}
Orbit of TianQin in ecliptic frame is given as follows ~\cite{Hu2018}:
\begin{widetext}
    \begin{equation}
        \begin{aligned}
            x_{n}(t) &=R_{1}\left(\cos \phi_{s} \sin \theta_{s} \sin \left(\alpha_{n}-\beta^{\prime}\right)+\cos \left(\alpha_{n}-\beta^{\prime}\right) \sin \phi_{s}\right)+R_{1} e_{1}\left[\frac{1}{2}\left(\cos 2\left(\alpha_{n}-\beta^{\prime}\right)-3\right) \sin \phi_{s}\right.\\
            &\left.+\cos \left(\alpha_{n}-\beta^{\prime}\right) \cos \phi_{s} \sin \theta_{s} \sin \left(\alpha_{n}-\beta^{\prime}\right)\right]+\frac{e_{1}^{2}}{4} R_{1} \sin \left(\alpha_{n}-\beta^{\prime}\right)\left[\left(3 \cos 2\left(\alpha_{n}-\beta^{\prime}\right)-1\right)\right.\\
            &\left.\times \cos \phi_{s} \sin \theta_{s}-6 \cos \left(\alpha_{n}-\beta^{\prime}\right) \sin \left(\alpha_{n}-\beta^{\prime}\right) \sin \phi_{s}\right]+R \cos (\alpha-\beta)+\frac{R e}{2}(\cos 2(\alpha-\beta)-3) \\
            &-\frac{3 R e^{2}}{2} \cos (\alpha-\beta) \sin ^{2}(\alpha-\beta), \\
            y_{n}(t) &=R_{1}\left(\sin \phi_{s} \sin \theta_{s} \sin \left(\alpha_{n}-\beta^{\prime}\right)-\cos \left(\alpha_{n}-\beta^{\prime}\right) \cos \phi_{s}\right)-R_{1} e_{1}\left[\frac{1}{2}\left(\cos 2\left(\alpha_{n}-\beta^{\prime}\right)-3\right) \cos \phi_{s}\right.\\
            &\left.-\cos \left(\alpha_{n}-\beta^{\prime}\right) \sin \phi_{s} \sin \theta_{s} \sin \left(\alpha_{n}-\beta^{\prime}\right)\right]+\frac{e_{1}^{2}}{4} R_{1} \sin \left(\alpha_{n}-\beta^{\prime}\right)\left[\left(3 \cos 2\left(\alpha_{n}-\beta^{\prime}\right)-1\right)\right.\\
            &\left.\times \sin \phi_{s} \sin \theta_{s}+6 \cos \left(\alpha_{n}-\beta^{\prime}\right) \sin \left(\alpha_{n}-\beta^{\prime}\right) \cos \phi_{s}\right]+R \sin (\alpha-\beta)+\frac{R e}{2} \sin 2(\alpha-\beta) \\
            &+\frac{R e^{2}}{4}(3 \cos 2(\alpha-\beta)-1) \sin (\alpha-\beta), \\
            z_{n}(t) &=-R_{1} \sin \left(\alpha_{n}-\beta^{\prime}\right) \cos \theta_{s}-R_{1} e_{1} \cos \left(\alpha_{n}-\beta^{\prime}\right) \sin \left(\alpha_{n}-\beta^{\prime}\right) \cos \theta_{s}, \\
            &-\frac{1}{4} e_{1}^{2} R_{1}\left(3 \cos 2\left(\alpha_{n}-\beta^{\prime}\right)-1\right) \sin \left(\alpha_{n}-\beta^{\prime}\right) \cos \theta_{s},
        \end{aligned}
    \end{equation}
\end{widetext}
where $R = 1$ AU and $e = 0.0167$ are the semi-major axis and the eccentricity of the geocenter orbit around the Sun; $R_1 = 1.0 \times 10^5$ km and $e_1$ are the semi-major axis and the eccentricity of the spacecraft orbit around the Earth. $\theta_s = -4.7^{\circ}, \phi_s = 120.5^{\circ} $ is the ecliptic coordinates of RX J0806.3+1527. $f_{\mathrm{m}}$ equals to $1/\mathrm{year}$ and $\alpha(t)=2 \pi f_{\mathrm{m}} t+\kappa_{0}$ is the mean ecliptic longitude of the geocenter in the heliocentric-ecliptic coordinate system. $\kappa_0$ is the mean ecliptic longitude measured from the vernal equinox at $t = 0$. $\beta$ is the longitude of the perihelion. $\alpha_n$ represents orbit phase of the n-th spacecraft. A specific introduction of the orbit can be found in ~\cite{Hu2018}.

\section{$t(f)$ in stationary phase approximation\label{App2}}
In stationary phase approximation, the relation $t(f)$ mentioned in Sec.~\ref{Sec3} takes the form~\cite{2004CQGra213771A,Arun_2005,2007CQGra24155V,zhao2018,Niu2020}

\begin{equation}
    \label{tf}
    t(f)=t_{c}-\frac{5}{256\left(G M_{c}\right)^{5 / 3}}( \pi f)^{-8 / 3} \sum_{i=0}^{7} \tau_{i}( \pi f G m)^{i / 3},
\end{equation}

with coefficients
\begin{widetext}
\begin{equation}
    \begin{aligned}
        \tau_{0} &=1, \\
        \tau_{1} &=0, \\
        \tau_{2} &=\frac{743}{252}+\frac{11}{3} \eta, \\
        \tau_{3} &=-\frac{32}{5} \pi, \\
        \tau_{4} &=\frac{3058673}{508032}+\frac{5429}{504} \eta+\frac{617}{72} \eta^{2}, \\
        \tau_{5} &=-\left(\frac{7729}{252}-\frac{13}{3} \eta\right) \pi, \\
        \tau_{6} &=-\frac{10052469856691}{23471078400}+\frac{128 \pi^{2}}{3}+\frac{6848 \gamma}{105}
        +\left(\frac{3147553127}{3048192}-\frac{451 \pi^{2}}{12}\right) \eta, \\
        &-\frac{15211}{1728} \eta^{2}+\frac{25565}{1296} \eta^{3}+\frac{3424}{105} \ln \left[16( \pi m f)^{2 / 3}\right], \\
        \tau_{7} &=\left(-\frac{15419335}{127008}-\frac{75703}{756} \eta+\frac{14809}{378} \eta^{2}\right) \pi,
    \end{aligned}
\end{equation}
\end{widetext}
where  $\gamma = 0.5772$ is the Euler-Mascheroni constant, $m$ is total mass $m_1+m_2$ of binary. $\eta = m_1m_2/(m_1+m_2)^2$ is the symmetric mass ratio and $M_c = \eta^{3/5}m$ is chirp mass.

Note that, the time-frequency relation $t(f)$ defined by Eq.~(\ref{tf}) is for the dominant term. For other modes, we have 
\begin{equation}
    t_{\ell m}(f) = t(2f/m).
\end{equation}

\bibliography{manuscript.bib}

\providecommand{\noopsort}[1]{}\providecommand{\singleletter}[1]{#1}%
\begin{thebibliography}{60}%
\makeatletter
\providecommand \@ifxundefined [1]{%
 \@ifx{#1\undefined}
}%
\providecommand \@ifnum [1]{%
 \ifnum #1\expandafter \@firstoftwo
 \else \expandafter \@secondoftwo
 \fi
}%
\providecommand \@ifx [1]{%
 \ifx #1\expandafter \@firstoftwo
 \else \expandafter \@secondoftwo
 \fi
}%
\providecommand \natexlab [1]{#1}%
\providecommand \enquote  [1]{``#1''}%
\providecommand \bibnamefont  [1]{#1}%
\providecommand \bibfnamefont [1]{#1}%
\providecommand \citenamefont [1]{#1}%
\providecommand \href@noop [0]{\@secondoftwo}%
\providecommand \href [0]{\begingroup \@sanitize@url \@href}%
\providecommand \@href[1]{\@@startlink{#1}\@@href}%
\providecommand \@@href[1]{\endgroup#1\@@endlink}%
\providecommand \@sanitize@url [0]{\catcode `\\12\catcode `\$12\catcode
  `\&12\catcode `\#12\catcode `\^12\catcode `\_12\catcode `\%12\relax}%
\providecommand \@@startlink[1]{}%
\providecommand \@@endlink[0]{}%
\providecommand \url  [0]{\begingroup\@sanitize@url \@url }%
\providecommand \@url [1]{\endgroup\@href {#1}{\urlprefix }}%
\providecommand \urlprefix  [0]{URL }%
\providecommand \Eprint [0]{\href }%
\providecommand \doibase [0]{https://doi.org/}%
\providecommand \selectlanguage [0]{\@gobble}%
\providecommand \bibinfo  [0]{\@secondoftwo}%
\providecommand \bibfield  [0]{\@secondoftwo}%
\providecommand \translation [1]{[#1]}%
\providecommand \BibitemOpen [0]{}%
\providecommand \bibitemStop [0]{}%
\providecommand \bibitemNoStop [0]{.\EOS\space}%
\providecommand \EOS [0]{\spacefactor3000\relax}%
\providecommand \BibitemShut  [1]{\csname bibitem#1\endcsname}%
\let\auto@bib@innerbib\@empty
\bibitem [{\citenamefont {Abbott}\ \emph
  {et~al.}(2019{\natexlab{a}})\citenamefont {Abbott}, \citenamefont {Abbott},
  \citenamefont {Abbott}, \citenamefont {Abraham}, \citenamefont {Acernese},
  \citenamefont {Ackley}, \citenamefont {Adams}, \citenamefont {Adhikari},
  \citenamefont {Adya}, \citenamefont {Affeldt},\ and\ \citenamefont
  {et~al.}}]{O1O2}%
  \BibitemOpen
  \bibfield  {author} {\bibinfo {author} {\bibfnamefont {B.}~\bibnamefont
  {Abbott}}, \bibinfo {author} {\bibfnamefont {R.}~\bibnamefont {Abbott}},
  \bibinfo {author} {\bibfnamefont {T.}~\bibnamefont {Abbott}}, \bibinfo
  {author} {\bibfnamefont {S.}~\bibnamefont {Abraham}}, \bibinfo {author}
  {\bibfnamefont {F.}~\bibnamefont {Acernese}}, \bibinfo {author}
  {\bibfnamefont {K.}~\bibnamefont {Ackley}}, \bibinfo {author} {\bibfnamefont
  {C.}~\bibnamefont {Adams}}, \bibinfo {author} {\bibfnamefont
  {R.}~\bibnamefont {Adhikari}}, \bibinfo {author} {\bibfnamefont
  {V.}~\bibnamefont {Adya}}, \bibinfo {author} {\bibfnamefont {C.}~\bibnamefont
  {Affeldt}},\ and\ \bibinfo {author} {\bibnamefont {et~al.}},\ }\bibfield
  {title} {\bibinfo {title} {Gwtc-1: A gravitational-wave transient catalog of
  compact binary mergers observed by ligo and virgo during the first and second
  observing runs},\ }\href {https://doi.org/10.1103/physrevx.9.031040}
  {\bibfield  {journal} {\bibinfo  {journal} {Physical Review X}\ }\textbf
  {\bibinfo {volume} {9}},\ \bibinfo {pages} {031040} (\bibinfo {year}
  {2019}{\natexlab{a}})},\ \Eprint {https://arxiv.org/abs/1811.12907}
  {arXiv:1811.12907 [astro-ph.HE]} \BibitemShut {NoStop}%
\bibitem [{\citenamefont {Hu}\ \emph {et~al.}(2018)\citenamefont {Hu},
  \citenamefont {Li}, \citenamefont {Wang}, \citenamefont {Feng}, \citenamefont
  {Zhou}, \citenamefont {Hu}, \citenamefont {Hu}, \citenamefont {Mei},\ and\
  \citenamefont {Shao}}]{Hu2018}%
  \BibitemOpen
  \bibfield  {author} {\bibinfo {author} {\bibfnamefont {X.~C.}\ \bibnamefont
  {Hu}}, \bibinfo {author} {\bibfnamefont {X.~H.}\ \bibnamefont {Li}}, \bibinfo
  {author} {\bibfnamefont {Y.}~\bibnamefont {Wang}}, \bibinfo {author}
  {\bibfnamefont {W.~F.}\ \bibnamefont {Feng}}, \bibinfo {author}
  {\bibfnamefont {M.~Y.}\ \bibnamefont {Zhou}}, \bibinfo {author}
  {\bibfnamefont {Y.~M.}\ \bibnamefont {Hu}}, \bibinfo {author} {\bibfnamefont
  {S.~C.}\ \bibnamefont {Hu}}, \bibinfo {author} {\bibfnamefont {J.~W.}\
  \bibnamefont {Mei}},\ and\ \bibinfo {author} {\bibfnamefont {C.~G.}\
  \bibnamefont {Shao}},\ }\bibfield  {title} {\bibinfo {title} {{Fundamentals
  of the orbit and response for TianQin}},\ }\href
  {https://doi.org/10.1088/1361-6382/aab52f} {\bibfield  {journal} {\bibinfo
  {journal} {Class. Quantum Gravity}\ }\textbf {\bibinfo {volume} {35}},\
  \bibinfo {pages} {1} (\bibinfo {year} {2018})},\ \Eprint
  {https://arxiv.org/abs/1803.03368} {arXiv:1803.03368} \BibitemShut {NoStop}%
\bibitem [{\citenamefont {Amaro-Seoane}\ \emph {et~al.}(2017)\citenamefont
  {Amaro-Seoane}, \citenamefont {Audley}, \citenamefont {Babak}, \citenamefont
  {Baker},\ and\ \citenamefont {et~al.}}]{Amaro-Seoane2017}%
  \BibitemOpen
  \bibfield  {author} {\bibinfo {author} {\bibfnamefont {P.}~\bibnamefont
  {Amaro-Seoane}}, \bibinfo {author} {\bibfnamefont {H.}~\bibnamefont
  {Audley}}, \bibinfo {author} {\bibfnamefont {S.}~\bibnamefont {Babak}},
  \bibinfo {author} {\bibnamefont {Baker}},\ and\ \bibinfo {author}
  {\bibnamefont {et~al.}},\ }\href {http://arxiv.org/abs/1702.00786} {\bibinfo
  {title} {{Laser Interferometer Space Antenna (LISA L3 mission proposal)}}}
  (\bibinfo {year} {2017}),\ \Eprint {https://arxiv.org/abs/1702.00786}
  {arXiv:1702.00786} \BibitemShut {NoStop}%
\bibitem [{\citenamefont {Liu}\ \emph {et~al.}(2020)\citenamefont {Liu},
  \citenamefont {Ruan},\ and\ \citenamefont {Guo}}]{Liu2020a}%
  \BibitemOpen
  \bibfield  {author} {\bibinfo {author} {\bibfnamefont {C.}~\bibnamefont
  {Liu}}, \bibinfo {author} {\bibfnamefont {W.-H.}\ \bibnamefont {Ruan}},\ and\
  \bibinfo {author} {\bibfnamefont {Z.-K.}\ \bibnamefont {Guo}},\ }\href
  {http://arxiv.org/abs/2006.04413} {\bibinfo {title} {{Constraining
  gravitational-wave polarizations with Taiji}}} (\bibinfo {year} {2020}),\
  \Eprint {https://arxiv.org/abs/2006.04413} {arXiv:2006.04413} \BibitemShut
  {NoStop}%
\bibitem [{\citenamefont {Singer}\ and\ \citenamefont
  {Price}(2016)}]{Singer2016}%
  \BibitemOpen
  \bibfield  {author} {\bibinfo {author} {\bibfnamefont {L.~P.}\ \bibnamefont
  {Singer}}\ and\ \bibinfo {author} {\bibfnamefont {L.~R.}\ \bibnamefont
  {Price}},\ }\bibfield  {title} {\bibinfo {title} {{Rapid Bayesian position
  reconstruction for gravitational-wave transients}},\ }\href
  {https://doi.org/10.1103/PhysRevD.93.024013} {\bibfield  {journal} {\bibinfo
  {journal} {Phys. Rev. D}\ }\textbf {\bibinfo {volume} {93}},\ \bibinfo
  {pages} {1} (\bibinfo {year} {2016})},\ \Eprint
  {https://arxiv.org/abs/1508.03634} {arXiv:1508.03634} \BibitemShut {NoStop}%
\bibitem [{\citenamefont {Abbott}\ \emph
  {et~al.}(2019{\natexlab{b}})\citenamefont {Abbott}, \citenamefont {Abbott},
  \citenamefont {Abbott}, \citenamefont {Acernese}, \citenamefont {Ackley},
  \citenamefont {Adams}, \citenamefont {Adams}, \citenamefont {Addesso},
  \citenamefont {Adhikari}, \citenamefont {Adya},\ and\ \citenamefont
  {et~al.}}]{GW170817}%
  \BibitemOpen
  \bibfield  {author} {\bibinfo {author} {\bibfnamefont {B.}~\bibnamefont
  {Abbott}}, \bibinfo {author} {\bibfnamefont {R.}~\bibnamefont {Abbott}},
  \bibinfo {author} {\bibfnamefont {T.}~\bibnamefont {Abbott}}, \bibinfo
  {author} {\bibfnamefont {F.}~\bibnamefont {Acernese}}, \bibinfo {author}
  {\bibfnamefont {K.}~\bibnamefont {Ackley}}, \bibinfo {author} {\bibfnamefont
  {C.}~\bibnamefont {Adams}}, \bibinfo {author} {\bibfnamefont
  {T.}~\bibnamefont {Adams}}, \bibinfo {author} {\bibfnamefont
  {P.}~\bibnamefont {Addesso}}, \bibinfo {author} {\bibfnamefont
  {R.}~\bibnamefont {Adhikari}}, \bibinfo {author} {\bibfnamefont
  {V.}~\bibnamefont {Adya}},\ and\ \bibinfo {author} {\bibnamefont {et~al.}},\
  }\bibfield  {title} {\bibinfo {title} {Properties of the binary neutron star
  merger gw170817},\ }\href {https://doi.org/10.1103/physrevx.9.011001}
  {\bibfield  {journal} {\bibinfo  {journal} {Physical Review X}\ }\textbf
  {\bibinfo {volume} {9}},\ \bibinfo {pages} {011001} (\bibinfo {year}
  {2019}{\natexlab{b}})}\BibitemShut {NoStop}%
\bibitem [{\citenamefont {Ruan}\ \emph {et~al.}(2021)\citenamefont {Ruan},
  \citenamefont {Liu}, \citenamefont {Guo}, \citenamefont {Wu},\ and\
  \citenamefont {Cai}}]{Ruan2019}%
  \BibitemOpen
  \bibfield  {author} {\bibinfo {author} {\bibfnamefont {W.-H.}\ \bibnamefont
  {Ruan}}, \bibinfo {author} {\bibfnamefont {C.}~\bibnamefont {Liu}}, \bibinfo
  {author} {\bibfnamefont {Z.-K.}\ \bibnamefont {Guo}}, \bibinfo {author}
  {\bibfnamefont {Y.-L.}\ \bibnamefont {Wu}},\ and\ \bibinfo {author}
  {\bibfnamefont {R.-G.}\ \bibnamefont {Cai}},\ }\bibfield  {title} {\bibinfo
  {title} {The lisa-taiji network: Precision localization of coalescing massive
  black hole binaries},\ }\href {https://doi.org/10.34133/2021/6014164}
  {\bibfield  {journal} {\bibinfo  {journal} {Research}\ }\textbf {\bibinfo
  {volume} {2021}},\ \bibinfo {pages} {1–7} (\bibinfo {year}
  {2021})}\BibitemShut {NoStop}%
\bibitem [{\citenamefont {Maggiore}(2007)}]{GW1}%
  \BibitemOpen
  \bibfield  {author} {\bibinfo {author} {\bibfnamefont {M.}~\bibnamefont
  {Maggiore}},\ }\href@noop {} {\emph {\bibinfo {title} {Gravitational Waves
  Vol. 1}}}\ (\bibinfo  {publisher} {Oxford University Press, Oxford,
  England},\ \bibinfo {year} {2007})\BibitemShut {NoStop}%
\bibitem [{\citenamefont {Maggiore}(2018)}]{GW2}%
  \BibitemOpen
  \bibfield  {author} {\bibinfo {author} {\bibfnamefont {M.}~\bibnamefont
  {Maggiore}},\ }\href@noop {} {\emph {\bibinfo {title} {Gravitational Waves,
  Vol. 2}}}\ (\bibinfo  {publisher} {Oxford University Press, Oxford,
  England},\ \bibinfo {year} {2018})\BibitemShut {NoStop}%
\bibitem [{\citenamefont {Lee}\ and\ \citenamefont {Yang}(1956)}]{Yang1956}%
  \BibitemOpen
  \bibfield  {author} {\bibinfo {author} {\bibfnamefont {T.~D.}\ \bibnamefont
  {Lee}}\ and\ \bibinfo {author} {\bibfnamefont {C.~N.}\ \bibnamefont {Yang}},\
  }\bibfield  {title} {\bibinfo {title} {Question of parity conservation in
  weak interactions},\ }\href {https://doi.org/10.1103/PhysRev.104.254}
  {\bibfield  {journal} {\bibinfo  {journal} {Phys. Rev.}\ }\textbf {\bibinfo
  {volume} {104}},\ \bibinfo {pages} {254} (\bibinfo {year}
  {1956})}\BibitemShut {NoStop}%
\bibitem [{\citenamefont {Alexander}\ and\ \citenamefont
  {Yunes}(2009)}]{Alexander_2009}%
  \BibitemOpen
  \bibfield  {author} {\bibinfo {author} {\bibfnamefont {S.}~\bibnamefont
  {Alexander}}\ and\ \bibinfo {author} {\bibfnamefont {N.}~\bibnamefont
  {Yunes}},\ }\bibfield  {title} {\bibinfo {title} {Chern–simons modified
  general relativity},\ }\href {https://doi.org/10.1016/j.physrep.2009.07.002}
  {\bibfield  {journal} {\bibinfo  {journal} {Physics Reports}\ }\textbf
  {\bibinfo {volume} {480}},\ \bibinfo {pages} {1–55} (\bibinfo {year}
  {2009})}\BibitemShut {NoStop}%
\bibitem [{\citenamefont {Campbell}\ \emph {et~al.}(1991)\citenamefont
  {Campbell}, \citenamefont {Duncan}, \citenamefont {Kaloper},\ and\
  \citenamefont {Olive}}]{CAMPBELL1991778}%
  \BibitemOpen
  \bibfield  {author} {\bibinfo {author} {\bibfnamefont {B.~A.}\ \bibnamefont
  {Campbell}}, \bibinfo {author} {\bibfnamefont {M.}~\bibnamefont {Duncan}},
  \bibinfo {author} {\bibfnamefont {N.}~\bibnamefont {Kaloper}},\ and\ \bibinfo
  {author} {\bibfnamefont {K.~A.}\ \bibnamefont {Olive}},\ }\bibfield  {title}
  {\bibinfo {title} {Gravitational dynamics with lorentz chern-simons terms},\
  }\href {https://doi.org/https://doi.org/10.1016/S0550-3213(05)80045-8}
  {\bibfield  {journal} {\bibinfo  {journal} {Nuclear Physics B}\ }\textbf
  {\bibinfo {volume} {351}},\ \bibinfo {pages} {778 } (\bibinfo {year}
  {1991})}\BibitemShut {NoStop}%
\bibitem [{\citenamefont {Campbell}\ \emph {et~al.}(1993)\citenamefont
  {Campbell}, \citenamefont {Kaloper}, \citenamefont {Madden},\ and\
  \citenamefont {Olive}}]{Campbell_1993}%
  \BibitemOpen
  \bibfield  {author} {\bibinfo {author} {\bibfnamefont {B.~A.}\ \bibnamefont
  {Campbell}}, \bibinfo {author} {\bibfnamefont {N.}~\bibnamefont {Kaloper}},
  \bibinfo {author} {\bibfnamefont {R.}~\bibnamefont {Madden}},\ and\ \bibinfo
  {author} {\bibfnamefont {K.~A.}\ \bibnamefont {Olive}},\ }\bibfield  {title}
  {\bibinfo {title} {Physical properties of four-dimensional superstring
  gravity black hole solutions},\ }\href
  {https://doi.org/10.1016/0550-3213(93)90620-5} {\bibfield  {journal}
  {\bibinfo  {journal} {Nuclear Physics B}\ }\textbf {\bibinfo {volume}
  {399}},\ \bibinfo {pages} {137–168} (\bibinfo {year} {1993})}\BibitemShut
  {NoStop}%
\bibitem [{\citenamefont {Zhao}\ \emph
  {et~al.}(2020{\natexlab{a}})\citenamefont {Zhao}, \citenamefont {Zhu},
  \citenamefont {Qiao},\ and\ \citenamefont {Wang}}]{Zhao_PVwaveform}%
  \BibitemOpen
  \bibfield  {author} {\bibinfo {author} {\bibfnamefont {W.}~\bibnamefont
  {Zhao}}, \bibinfo {author} {\bibfnamefont {T.}~\bibnamefont {Zhu}}, \bibinfo
  {author} {\bibfnamefont {J.}~\bibnamefont {Qiao}},\ and\ \bibinfo {author}
  {\bibfnamefont {A.}~\bibnamefont {Wang}},\ }\bibfield  {title} {\bibinfo
  {title} {{Waveform of gravitational waves in the general parity-violating
  gravities}},\ }\href {https://doi.org/10.1103/PhysRevD.101.024002} {\bibfield
   {journal} {\bibinfo  {journal} {Phys. Rev. D}\ }\textbf {\bibinfo {volume}
  {101}},\ \bibinfo {pages} {024002} (\bibinfo {year} {2020}{\natexlab{a}})},\
  \Eprint {https://arxiv.org/abs/1909.10887} {arXiv:1909.10887} \BibitemShut
  {NoStop}%
\bibitem [{\citenamefont {Qiao}\ \emph {et~al.}(2019)\citenamefont {Qiao},
  \citenamefont {Zhu}, \citenamefont {Zhao},\ and\ \citenamefont
  {Wang}}]{Qiao2019}%
  \BibitemOpen
  \bibfield  {author} {\bibinfo {author} {\bibfnamefont {J.}~\bibnamefont
  {Qiao}}, \bibinfo {author} {\bibfnamefont {T.}~\bibnamefont {Zhu}}, \bibinfo
  {author} {\bibfnamefont {W.}~\bibnamefont {Zhao}},\ and\ \bibinfo {author}
  {\bibfnamefont {A.}~\bibnamefont {Wang}},\ }\bibfield  {title} {\bibinfo
  {title} {Waveform of gravitational waves in the ghost-free parity-violating
  gravities},\ }\href {https://doi.org/10.1103/PhysRevD.100.124058} {\bibfield
  {journal} {\bibinfo  {journal} {Phys. Rev. D}\ }\textbf {\bibinfo {volume}
  {100}},\ \bibinfo {pages} {124058} (\bibinfo {year} {2019})},\ \Eprint
  {https://arxiv.org/abs/1909.03815} {arXiv:1909.03815} \BibitemShut {NoStop}%
\bibitem [{\citenamefont {{Wang}}\ \emph {et~al.}(2013)\citenamefont {{Wang}},
  \citenamefont {{Wu}}, \citenamefont {{Zhao}},\ and\ \citenamefont
  {{Zhu}}}]{wang2013}%
  \BibitemOpen
  \bibfield  {author} {\bibinfo {author} {\bibfnamefont {A.}~\bibnamefont
  {{Wang}}}, \bibinfo {author} {\bibfnamefont {Q.}~\bibnamefont {{Wu}}},
  \bibinfo {author} {\bibfnamefont {W.}~\bibnamefont {{Zhao}}},\ and\ \bibinfo
  {author} {\bibfnamefont {T.}~\bibnamefont {{Zhu}}},\ }\bibfield  {title}
  {\bibinfo {title} {{Polarizing primordial gravitational waves by parity
  violation}},\ }\href {https://doi.org/10.1103/PhysRevD.87.103512} {\bibfield
  {journal} {\bibinfo  {journal} {\prd}\ }\textbf {\bibinfo {volume} {87}},\
  \bibinfo {eid} {103512} (\bibinfo {year} {2013})},\ \Eprint
  {https://arxiv.org/abs/1208.5490} {arXiv:1208.5490 [astro-ph.CO]}
  \BibitemShut {NoStop}%
\bibitem [{\citenamefont {{Zhu}}\ \emph {et~al.}(2013)\citenamefont {{Zhu}},
  \citenamefont {{Zhao}}, \citenamefont {{Huang}}, \citenamefont {{Wang}},\
  and\ \citenamefont {{Wu}}}]{zhu2013}%
  \BibitemOpen
  \bibfield  {author} {\bibinfo {author} {\bibfnamefont {T.}~\bibnamefont
  {{Zhu}}}, \bibinfo {author} {\bibfnamefont {W.}~\bibnamefont {{Zhao}}},
  \bibinfo {author} {\bibfnamefont {Y.}~\bibnamefont {{Huang}}}, \bibinfo
  {author} {\bibfnamefont {A.}~\bibnamefont {{Wang}}},\ and\ \bibinfo {author}
  {\bibfnamefont {Q.}~\bibnamefont {{Wu}}},\ }\bibfield  {title} {\bibinfo
  {title} {{Effects of parity violation on non-Gaussianity of primordial
  gravitational waves in Ho{\v{r}}ava-Lifshitz gravity}},\ }\href
  {https://doi.org/10.1103/PhysRevD.88.063508} {\bibfield  {journal} {\bibinfo
  {journal} {\prd}\ }\textbf {\bibinfo {volume} {88}},\ \bibinfo {eid} {063508}
  (\bibinfo {year} {2013})},\ \Eprint {https://arxiv.org/abs/1305.0600}
  {arXiv:1305.0600 [hep-th]} \BibitemShut {NoStop}%
\bibitem [{\citenamefont {{Qiao}}\ \emph {et~al.}(2020)\citenamefont {{Qiao}},
  \citenamefont {{Zhu}}, \citenamefont {{Zhao}},\ and\ \citenamefont
  {{Wang}}}]{qiao2020}%
  \BibitemOpen
  \bibfield  {author} {\bibinfo {author} {\bibfnamefont {J.}~\bibnamefont
  {{Qiao}}}, \bibinfo {author} {\bibfnamefont {T.}~\bibnamefont {{Zhu}}},
  \bibinfo {author} {\bibfnamefont {W.}~\bibnamefont {{Zhao}}},\ and\ \bibinfo
  {author} {\bibfnamefont {A.}~\bibnamefont {{Wang}}},\ }\bibfield  {title}
  {\bibinfo {title} {{Polarized primordial gravitational waves in the
  ghost-free parity-violating gravity}},\ }\href
  {https://doi.org/10.1103/PhysRevD.101.043528} {\bibfield  {journal} {\bibinfo
   {journal} {\prd}\ }\textbf {\bibinfo {volume} {101}},\ \bibinfo {eid}
  {043528} (\bibinfo {year} {2020})},\ \Eprint
  {https://arxiv.org/abs/1911.01580} {arXiv:1911.01580 [astro-ph.CO]}
  \BibitemShut {NoStop}%
\bibitem [{\citenamefont {Wang}\ \emph {et~al.}(2021)\citenamefont {Wang},
  \citenamefont {Niu}, \citenamefont {Zhu},\ and\ \citenamefont
  {Zhao}}]{Wang2020}%
  \BibitemOpen
  \bibfield  {author} {\bibinfo {author} {\bibfnamefont {Y.-F.}\ \bibnamefont
  {Wang}}, \bibinfo {author} {\bibfnamefont {R.}~\bibnamefont {Niu}}, \bibinfo
  {author} {\bibfnamefont {T.}~\bibnamefont {Zhu}},\ and\ \bibinfo {author}
  {\bibfnamefont {W.}~\bibnamefont {Zhao}},\ }\bibfield  {title} {\bibinfo
  {title} {{Gravitational-Wave Implications for the Parity Symmetry of Gravity
  in the High Energy Region}},\ }\href
  {https://doi.org/10.3847/1538-4357/abd7a6} {\bibfield  {journal} {\bibinfo
  {journal} {Astrophys. J.}\ }\textbf {\bibinfo {volume} {908}},\ \bibinfo
  {pages} {58} (\bibinfo {year} {2021})},\ \Eprint
  {https://arxiv.org/abs/2002.05668} {arXiv:2002.05668} \BibitemShut {NoStop}%
\bibitem [{\citenamefont {Kostelecký}\ and\ \citenamefont
  {Mewes}(2016)}]{Kosteleck__2016}%
  \BibitemOpen
  \bibfield  {author} {\bibinfo {author} {\bibfnamefont {V.~A.}\ \bibnamefont
  {Kostelecký}}\ and\ \bibinfo {author} {\bibfnamefont {M.}~\bibnamefont
  {Mewes}},\ }\bibfield  {title} {\bibinfo {title} {Testing local lorentz
  invariance with gravitational waves},\ }\href
  {https://doi.org/10.1016/j.physletb.2016.04.040} {\bibfield  {journal}
  {\bibinfo  {journal} {Physics Letters B}\ }\textbf {\bibinfo {volume}
  {757}},\ \bibinfo {pages} {510–514} (\bibinfo {year} {2016})},\ \Eprint
  {https://arxiv.org/abs/1602.04782} {arXiv:1602.04782 [gr-qc]} \BibitemShut
  {NoStop}%
\bibitem [{\citenamefont {Yoshida}\ and\ \citenamefont
  {Soda}(2018)}]{Yoshida_2018}%
  \BibitemOpen
  \bibfield  {author} {\bibinfo {author} {\bibfnamefont {D.}~\bibnamefont
  {Yoshida}}\ and\ \bibinfo {author} {\bibfnamefont {J.}~\bibnamefont {Soda}},\
  }\bibfield  {title} {\bibinfo {title} {Exploring the string axiverse and
  parity violation in gravity with gravitational waves},\ }\href
  {https://doi.org/10.1142/s0218271818500967} {\bibfield  {journal} {\bibinfo
  {journal} {International Journal of Modern Physics D}\ }\textbf {\bibinfo
  {volume} {27}},\ \bibinfo {pages} {1850096} (\bibinfo {year} {2018})},\
  \Eprint {https://arxiv.org/abs/1708.09592} {arXiv:1708.09592 [gr-qc]}
  \BibitemShut {NoStop}%
\bibitem [{\citenamefont {Yagi}\ and\ \citenamefont {Yang}(2018)}]{Yagi_2018}%
  \BibitemOpen
  \bibfield  {author} {\bibinfo {author} {\bibfnamefont {K.}~\bibnamefont
  {Yagi}}\ and\ \bibinfo {author} {\bibfnamefont {H.}~\bibnamefont {Yang}},\
  }\bibfield  {title} {\bibinfo {title} {Probing gravitational parity violation
  with gravitational waves from stellar-mass black hole binaries},\ }\href
  {https://doi.org/10.1103/physrevd.97.104018} {\bibfield  {journal} {\bibinfo
  {journal} {Physical Review D}\ }\textbf {\bibinfo {volume} {97}},\ \bibinfo
  {pages} {104018} (\bibinfo {year} {2018})},\ \Eprint
  {https://arxiv.org/abs/1712.00682} {arXiv:1712.00682 [gr-qc]} \BibitemShut
  {NoStop}%
\bibitem [{\citenamefont {Alexander}\ and\ \citenamefont
  {Yunes}(2018)}]{Alexander_2018}%
  \BibitemOpen
  \bibfield  {author} {\bibinfo {author} {\bibfnamefont {S.~H.}\ \bibnamefont
  {Alexander}}\ and\ \bibinfo {author} {\bibfnamefont {N.}~\bibnamefont
  {Yunes}},\ }\bibfield  {title} {\bibinfo {title} {Gravitational wave probes
  of parity violation in compact binary coalescences},\ }\href
  {https://doi.org/10.1103/physrevd.97.064033} {\bibfield  {journal} {\bibinfo
  {journal} {Physical Review D}\ }\textbf {\bibinfo {volume} {97}},\ \bibinfo
  {pages} {064003} (\bibinfo {year} {2018})},\ \Eprint
  {https://arxiv.org/abs/1712.01853} {arXiv:1712.01853 [gr-qc]} \BibitemShut
  {NoStop}%
\bibitem [{\citenamefont {Silva}\ \emph {et~al.}(2020)\citenamefont {Silva},
  \citenamefont {Holgado}, \citenamefont {Cárdenas-Avendaño},\ and\
  \citenamefont {Yunes}}]{silva2020astrophysical}%
  \BibitemOpen
  \bibfield  {author} {\bibinfo {author} {\bibfnamefont {H.~O.}\ \bibnamefont
  {Silva}}, \bibinfo {author} {\bibfnamefont {A.~M.}\ \bibnamefont {Holgado}},
  \bibinfo {author} {\bibfnamefont {A.}~\bibnamefont {Cárdenas-Avendaño}},\
  and\ \bibinfo {author} {\bibfnamefont {N.}~\bibnamefont {Yunes}},\
  }\href@noop {} {\bibinfo {title} {Astrophysical and theoretical physics
  implications from multimessenger neutron star observations}} (\bibinfo {year}
  {2020}),\ \Eprint {https://arxiv.org/abs/2004.01253} {arXiv:2004.01253
  [gr-qc]} \BibitemShut {NoStop}%
\bibitem [{\citenamefont {Shao}(2020)}]{Shaolj_2020}%
  \BibitemOpen
  \bibfield  {author} {\bibinfo {author} {\bibfnamefont {L.}~\bibnamefont
  {Shao}},\ }\bibfield  {title} {\bibinfo {title} {Combined search for
  anisotropic birefringence in the gravitational-wave transient catalog
  gwtc-1},\ }\href {https://doi.org/10.1103/physrevd.101.104019} {\bibfield
  {journal} {\bibinfo  {journal} {Physical Review D}\ }\textbf {\bibinfo
  {volume} {101}},\ \bibinfo {pages} {104019} (\bibinfo {year}
  {2020})}\BibitemShut {NoStop}%
\bibitem [{\citenamefont {Creminelli}\ \emph {et~al.}(2014)\citenamefont
  {Creminelli}, \citenamefont {Gleyzes}, \citenamefont {Nore\~na},\ and\
  \citenamefont {Vernizzi}}]{PhysRevLett113231301}%
  \BibitemOpen
  \bibfield  {author} {\bibinfo {author} {\bibfnamefont {P.}~\bibnamefont
  {Creminelli}}, \bibinfo {author} {\bibfnamefont {J.}~\bibnamefont {Gleyzes}},
  \bibinfo {author} {\bibfnamefont {J.}~\bibnamefont {Nore\~na}},\ and\
  \bibinfo {author} {\bibfnamefont {F.}~\bibnamefont {Vernizzi}},\ }\bibfield
  {title} {\bibinfo {title} {Resilience of the standard predictions for
  primordial tensor modes},\ }\href
  {https://doi.org/10.1103/PhysRevLett.113.231301} {\bibfield  {journal}
  {\bibinfo  {journal} {Phys. Rev. Lett.}\ }\textbf {\bibinfo {volume} {113}},\
  \bibinfo {pages} {231301} (\bibinfo {year} {2014})}\BibitemShut {NoStop}%
\bibitem [{\citenamefont {Misner}\ \emph {et~al.}(1973)\citenamefont {Misner},
  \citenamefont {Thorne},\ and\ \citenamefont {Wheeler}}]{MTW}%
  \BibitemOpen
  \bibfield  {author} {\bibinfo {author} {\bibfnamefont {C.~W.}\ \bibnamefont
  {Misner}}, \bibinfo {author} {\bibfnamefont {K.}~\bibnamefont {Thorne}},\
  and\ \bibinfo {author} {\bibfnamefont {J.}~\bibnamefont {Wheeler}},\
  }\href@noop {} {\emph {\bibinfo {title} {{Gravitation}}}}\ (\bibinfo
  {publisher} {W. H. Freeman},\ \bibinfo {address} {San Francisco},\ \bibinfo
  {year} {1973})\BibitemShut {NoStop}%
\bibitem [{\citenamefont {Adam}\ \emph {et~al.}(2016)\citenamefont {Adam},
  \citenamefont {Ade}, \citenamefont {Aghanim}, \citenamefont {Akrami},
  \citenamefont {Alves}, \citenamefont {Argüeso}, \citenamefont {Arnaud},
  \citenamefont {Arroja}, \citenamefont {Ashdown},\ and\ \citenamefont
  {et~al.}}]{Planck1}%
  \BibitemOpen
  \bibfield  {author} {\bibinfo {author} {\bibfnamefont {R.}~\bibnamefont
  {Adam}}, \bibinfo {author} {\bibfnamefont {P.~A.~R.}\ \bibnamefont {Ade}},
  \bibinfo {author} {\bibfnamefont {N.}~\bibnamefont {Aghanim}}, \bibinfo
  {author} {\bibfnamefont {Y.}~\bibnamefont {Akrami}}, \bibinfo {author}
  {\bibfnamefont {M.~I.~R.}\ \bibnamefont {Alves}}, \bibinfo {author}
  {\bibfnamefont {F.}~\bibnamefont {Argüeso}}, \bibinfo {author}
  {\bibfnamefont {M.}~\bibnamefont {Arnaud}}, \bibinfo {author} {\bibfnamefont
  {F.}~\bibnamefont {Arroja}}, \bibinfo {author} {\bibfnamefont
  {M.}~\bibnamefont {Ashdown}},\ and\ \bibinfo {author} {\bibnamefont
  {et~al.}},\ }\bibfield  {title} {\bibinfo {title} {Planck 2015 results. i.
  overview of products and scientific results},\ }\href
  {https://doi.org/10.1051/0004-6361/201527101} {\bibfield  {journal} {\bibinfo
   {journal} {Astronomy and Astrophysics}\ }\textbf {\bibinfo {volume} {594}},\
  \bibinfo {pages} {A1} (\bibinfo {year} {2016})}\BibitemShut {NoStop}%
\bibitem [{\citenamefont {Ade}\ \emph {et~al.}(2016)\citenamefont {Ade},
  \citenamefont {Aghanim}, \citenamefont {Arnaud}, \citenamefont {Ashdown},
  \citenamefont {Aumont}, \citenamefont {Baccigalupi}, \citenamefont {Banday},
  \citenamefont {Barreiro}, \citenamefont {Bartlett},\ and\ \citenamefont
  {et~al.}}]{Planck2}%
  \BibitemOpen
  \bibfield  {author} {\bibinfo {author} {\bibfnamefont {P.~A.~R.}\
  \bibnamefont {Ade}}, \bibinfo {author} {\bibfnamefont {N.}~\bibnamefont
  {Aghanim}}, \bibinfo {author} {\bibfnamefont {M.}~\bibnamefont {Arnaud}},
  \bibinfo {author} {\bibfnamefont {M.}~\bibnamefont {Ashdown}}, \bibinfo
  {author} {\bibfnamefont {J.}~\bibnamefont {Aumont}}, \bibinfo {author}
  {\bibfnamefont {C.}~\bibnamefont {Baccigalupi}}, \bibinfo {author}
  {\bibfnamefont {A.~J.}\ \bibnamefont {Banday}}, \bibinfo {author}
  {\bibfnamefont {R.~B.}\ \bibnamefont {Barreiro}}, \bibinfo {author}
  {\bibfnamefont {J.~G.}\ \bibnamefont {Bartlett}},\ and\ \bibinfo {author}
  {\bibnamefont {et~al.}},\ }\bibfield  {title} {\bibinfo {title} {Planck 2015
  results. xiii. cosmological parameters},\ }\href
  {https://doi.org/10.1051/0004-6361/201525830} {\bibfield  {journal} {\bibinfo
   {journal} {Astronomy and Astrophysics}\ }\textbf {\bibinfo {volume} {594}},\
  \bibinfo {pages} {A13} (\bibinfo {year} {2016})}\BibitemShut {NoStop}%
\bibitem [{\citenamefont {Belgacem}\ \emph {et~al.}(2019)\citenamefont
  {Belgacem}, \citenamefont {Calcagni}, \citenamefont {Crisostomi},
  \citenamefont {Dalang}, \citenamefont {Dirian}, \citenamefont {Ezquiaga},
  \citenamefont {Fasiello}, \citenamefont {Foffa}, \citenamefont {Ganz},
  \citenamefont {Garc{\'{i}}a-Bellido}, \citenamefont {Lombriser},
  \citenamefont {Maggiore}, \citenamefont {Tamanini}, \citenamefont {Tasinato},
  \citenamefont {Zumalac{\'{a}}rregui}, \citenamefont {Barausse}, \citenamefont
  {Bartolo}, \citenamefont {Bertacca}, \citenamefont {Klein}, \citenamefont
  {Matarrese},\ and\ \citenamefont {Sakellariadou}}]{Belgacem2019}%
  \BibitemOpen
  \bibfield  {author} {\bibinfo {author} {\bibfnamefont {E.}~\bibnamefont
  {Belgacem}}, \bibinfo {author} {\bibfnamefont {G.}~\bibnamefont {Calcagni}},
  \bibinfo {author} {\bibfnamefont {M.}~\bibnamefont {Crisostomi}}, \bibinfo
  {author} {\bibfnamefont {C.}~\bibnamefont {Dalang}}, \bibinfo {author}
  {\bibfnamefont {Y.}~\bibnamefont {Dirian}}, \bibinfo {author} {\bibfnamefont
  {J.~M.}\ \bibnamefont {Ezquiaga}}, \bibinfo {author} {\bibfnamefont
  {M.}~\bibnamefont {Fasiello}}, \bibinfo {author} {\bibfnamefont
  {S.}~\bibnamefont {Foffa}}, \bibinfo {author} {\bibfnamefont
  {A.}~\bibnamefont {Ganz}}, \bibinfo {author} {\bibfnamefont {J.}~\bibnamefont
  {Garc{\'{i}}a-Bellido}}, \bibinfo {author} {\bibfnamefont {L.}~\bibnamefont
  {Lombriser}}, \bibinfo {author} {\bibfnamefont {M.}~\bibnamefont {Maggiore}},
  \bibinfo {author} {\bibfnamefont {N.}~\bibnamefont {Tamanini}}, \bibinfo
  {author} {\bibfnamefont {G.}~\bibnamefont {Tasinato}}, \bibinfo {author}
  {\bibfnamefont {M.}~\bibnamefont {Zumalac{\'{a}}rregui}}, \bibinfo {author}
  {\bibfnamefont {E.}~\bibnamefont {Barausse}}, \bibinfo {author}
  {\bibfnamefont {N.}~\bibnamefont {Bartolo}}, \bibinfo {author} {\bibfnamefont
  {D.}~\bibnamefont {Bertacca}}, \bibinfo {author} {\bibfnamefont
  {A.}~\bibnamefont {Klein}}, \bibinfo {author} {\bibfnamefont
  {S.}~\bibnamefont {Matarrese}},\ and\ \bibinfo {author} {\bibfnamefont
  {M.}~\bibnamefont {Sakellariadou}},\ }\bibfield  {title} {\bibinfo {title}
  {{Testing modified gravity at cosmological distances with LISA standard
  sirens}},\ }\href {https://doi.org/10.1088/1475-7516/2019/07/024} {\bibfield
  {journal} {\bibinfo  {journal} {J. Cosmol. Astropart. Phys.}\ }\textbf
  {\bibinfo {volume} {2019}}\bibfield  {number} {\bibinfo  {number} { (7)},\
  \bibinfo {pages} {0}},\ }\Eprint {https://arxiv.org/abs/1906.01593}
  {arXiv:1906.01593} \BibitemShut {NoStop}%
\bibitem [{\citenamefont {Huang}\ \emph {et~al.}(2020)\citenamefont {Huang},
  \citenamefont {Hu}, \citenamefont {Korol}, \citenamefont {Li}, \citenamefont
  {Liang}, \citenamefont {Lu}, \citenamefont {Wang}, \citenamefont {Yu},\ and\
  \citenamefont {Mei}}]{Huang2020}%
  \BibitemOpen
  \bibfield  {author} {\bibinfo {author} {\bibfnamefont {S.-J.}\ \bibnamefont
  {Huang}}, \bibinfo {author} {\bibfnamefont {Y.-M.}\ \bibnamefont {Hu}},
  \bibinfo {author} {\bibfnamefont {V.}~\bibnamefont {Korol}}, \bibinfo
  {author} {\bibfnamefont {P.-C.}\ \bibnamefont {Li}}, \bibinfo {author}
  {\bibfnamefont {Z.-C.}\ \bibnamefont {Liang}}, \bibinfo {author}
  {\bibfnamefont {Y.}~\bibnamefont {Lu}}, \bibinfo {author} {\bibfnamefont
  {H.-T.}\ \bibnamefont {Wang}}, \bibinfo {author} {\bibfnamefont
  {S.}~\bibnamefont {Yu}},\ and\ \bibinfo {author} {\bibfnamefont
  {J.}~\bibnamefont {Mei}},\ }\href {http://arxiv.org/abs/2005.07889} {\bibinfo
  {title} {{Science with the TianQin Observatory: Preliminary Results on
  Galactic Double White Dwarf Binaries}}} (\bibinfo {year} {2020}),\ \Eprint
  {https://arxiv.org/abs/2005.07889} {arXiv:2005.07889} \BibitemShut {NoStop}%
\bibitem [{\citenamefont {Liu}\ \emph {et~al.}()\citenamefont {Liu},
  \citenamefont {Zhang}, \citenamefont {Gong}, \citenamefont {Wang},\ and\
  \citenamefont {Wang}}]{Liu2020}%
  \BibitemOpen
  \bibfield  {author} {\bibinfo {author} {\bibfnamefont {H.}~\bibnamefont
  {Liu}}, \bibinfo {author} {\bibfnamefont {C.}~\bibnamefont {Zhang}}, \bibinfo
  {author} {\bibfnamefont {Y.}~\bibnamefont {Gong}}, \bibinfo {author}
  {\bibfnamefont {B.}~\bibnamefont {Wang}},\ and\ \bibinfo {author}
  {\bibfnamefont {A.}~\bibnamefont {Wang}},\ }\bibfield  {title} {\bibinfo
  {title} {{Exploring non-singular black holes in gravitational
  perturbations}},\ }\href {http://arxiv.org/abs/2002.06360} {\ }\Eprint
  {https://arxiv.org/abs/2002.06360} {arXiv:2002.06360} \BibitemShut {NoStop}%
\bibitem [{\citenamefont {Cutler}(1998)}]{Cutler1998}%
  \BibitemOpen
  \bibfield  {author} {\bibinfo {author} {\bibfnamefont {C.}~\bibnamefont
  {Cutler}},\ }\bibfield  {title} {\bibinfo {title} {{Angular resolution of the
  LISA gravitational wave detector}},\ }\href
  {https://doi.org/10.1103/PhysRevD.57.7089} {\bibfield  {journal} {\bibinfo
  {journal} {Phys. Rev. D - Part. Fields, Gravit. Cosmol.}\ }\textbf {\bibinfo
  {volume} {57}},\ \bibinfo {pages} {7089} (\bibinfo {year} {1998})},\ \Eprint
  {https://arxiv.org/abs/9703068} {arXiv:9703068 [gr-qc]} \BibitemShut
  {NoStop}%
\bibitem [{\citenamefont {Cornish}\ and\ \citenamefont
  {Larson}(2001)}]{Cornish_2001}%
  \BibitemOpen
  \bibfield  {author} {\bibinfo {author} {\bibfnamefont {N.~J.}\ \bibnamefont
  {Cornish}}\ and\ \bibinfo {author} {\bibfnamefont {S.~L.}\ \bibnamefont
  {Larson}},\ }\bibfield  {title} {\bibinfo {title} {Space missions to detect
  the cosmic gravitational-wave background},\ }\href
  {https://doi.org/10.1088/0264-9381/18/17/308} {\bibfield  {journal} {\bibinfo
   {journal} {Classical and Quantum Gravity}\ }\textbf {\bibinfo {volume}
  {18}},\ \bibinfo {pages} {3473} (\bibinfo {year} {2001})}\BibitemShut
  {NoStop}%
\bibitem [{\citenamefont {Liang}\ \emph {et~al.}(2019)\citenamefont {Liang},
  \citenamefont {Gong}, \citenamefont {Weinstein}, \citenamefont {Zhang},\ and\
  \citenamefont {Zhang}}]{Liang2019}%
  \BibitemOpen
  \bibfield  {author} {\bibinfo {author} {\bibfnamefont {D.}~\bibnamefont
  {Liang}}, \bibinfo {author} {\bibfnamefont {Y.}~\bibnamefont {Gong}},
  \bibinfo {author} {\bibfnamefont {A.~J.}\ \bibnamefont {Weinstein}}, \bibinfo
  {author} {\bibfnamefont {C.}~\bibnamefont {Zhang}},\ and\ \bibinfo {author}
  {\bibfnamefont {C.}~\bibnamefont {Zhang}},\ }\bibfield  {title} {\bibinfo
  {title} {{Frequency response of space-based interferometric
  gravitational-wave detectors}},\ }\href
  {https://doi.org/10.1103/PhysRevD.99.104027} {\bibfield  {journal} {\bibinfo
  {journal} {Phys. Rev. D}\ }\textbf {\bibinfo {volume} {99}},\ \bibinfo
  {pages} {104027} (\bibinfo {year} {2019})},\ \Eprint
  {https://arxiv.org/abs/1901.09624} {arXiv:1901.09624} \BibitemShut {NoStop}%
\bibitem [{\citenamefont {Marsat}\ \emph {et~al.}()\citenamefont {Marsat},
  \citenamefont {Baker},\ and\ \citenamefont {Canton}}]{Marsat2020}%
  \BibitemOpen
  \bibfield  {author} {\bibinfo {author} {\bibfnamefont {S.}~\bibnamefont
  {Marsat}}, \bibinfo {author} {\bibfnamefont {J.~G.}\ \bibnamefont {Baker}},\
  and\ \bibinfo {author} {\bibfnamefont {T.~D.}\ \bibnamefont {Canton}},\
  }\href {http://arxiv.org/abs/2003.00357} {\bibinfo {title} {{Exploring the
  Bayesian parameter estimation of binary black holes with LISA}}},\ \Eprint
  {https://arxiv.org/abs/2003.00357} {arXiv:2003.00357} \BibitemShut {NoStop}%
\bibitem [{\citenamefont {Thrane}\ and\ \citenamefont
  {Talbot}(2019)}]{Thrane2019}%
  \BibitemOpen
  \bibfield  {author} {\bibinfo {author} {\bibfnamefont {E.}~\bibnamefont
  {Thrane}}\ and\ \bibinfo {author} {\bibfnamefont {C.}~\bibnamefont
  {Talbot}},\ }\bibfield  {title} {\bibinfo {title} {{An introduction to
  Bayesian inference in gravitational-wave astronomy: Parameter estimation,
  model selection, and hierarchical models}},\ }\bibfield  {journal} {\bibinfo
  {journal} {Publ. Astron. Soc. Aust.}\ }\textbf {\bibinfo {volume} {36}},\
  \href {https://doi.org/10.1017/pasa.2019.2} {10.1017/pasa.2019.2} (\bibinfo
  {year} {2019}),\ \Eprint {https://arxiv.org/abs/1809.02293}
  {arXiv:1809.02293} \BibitemShut {NoStop}%
\bibitem [{\citenamefont {Feroz}\ \emph {et~al.}(2009)\citenamefont {Feroz},
  \citenamefont {Hobson},\ and\ \citenamefont {Bridges}}]{Multinest}%
  \BibitemOpen
  \bibfield  {author} {\bibinfo {author} {\bibfnamefont {F.}~\bibnamefont
  {Feroz}}, \bibinfo {author} {\bibfnamefont {M.~P.}\ \bibnamefont {Hobson}},\
  and\ \bibinfo {author} {\bibfnamefont {M.}~\bibnamefont {Bridges}},\
  }\bibfield  {title} {\bibinfo {title} {{MultiNest: An efficient and robust
  Bayesian inference tool for cosmology and particle physics}},\ }\href
  {https://doi.org/10.1111/j.1365-2966.2009.14548.x} {\bibfield  {journal}
  {\bibinfo  {journal} {Mon. Not. R. Astron. Soc.}\ }\textbf {\bibinfo {volume}
  {398}},\ \bibinfo {pages} {1601} (\bibinfo {year} {2009})},\ \Eprint
  {https://arxiv.org/abs/0809.3437} {arXiv:0809.3437} \BibitemShut {NoStop}%
\bibitem [{\citenamefont {Skilling}(2004)}]{NestedSampling}%
  \BibitemOpen
  \bibfield  {author} {\bibinfo {author} {\bibfnamefont {J.}~\bibnamefont
  {Skilling}},\ }\bibfield  {title} {\bibinfo {title} {Nested sampling},\
  }\href {https://doi.org/10.1063/1.1835238} {\bibfield  {journal} {\bibinfo
  {journal} {AIP Conference Proceedings}\ }\textbf {\bibinfo {volume} {735}},\
  \bibinfo {pages} {395} (\bibinfo {year} {2004})},\ \Eprint
  {https://arxiv.org/abs/https://aip.scitation.org/doi/pdf/10.1063/1.1835238}
  {https://aip.scitation.org/doi/pdf/10.1063/1.1835238} \BibitemShut {NoStop}%
\bibitem [{\citenamefont {Veitch}\ \emph {et~al.}(2015)\citenamefont {Veitch},
  \citenamefont {Raymond},\ and\ \citenamefont {et~al.}}]{LALInference}%
  \BibitemOpen
  \bibfield  {author} {\bibinfo {author} {\bibfnamefont {J.}~\bibnamefont
  {Veitch}}, \bibinfo {author} {\bibfnamefont {V.}~\bibnamefont {Raymond}},\
  and\ \bibinfo {author} {\bibnamefont {et~al.}},\ }\bibfield  {title}
  {\bibinfo {title} {Parameter estimation for compact binaries with
  ground-based gravitational-wave observations using the lalinference software
  library},\ }\href {https://doi.org/10.1103/PhysRevD.91.042003} {\bibfield
  {journal} {\bibinfo  {journal} {Phys. Rev. D}\ }\textbf {\bibinfo {volume}
  {91}},\ \bibinfo {pages} {042003} (\bibinfo {year} {2015})}\BibitemShut
  {NoStop}%
\bibitem [{\citenamefont {Biwer}\ \emph {et~al.}(2019)\citenamefont {Biwer},
  \citenamefont {Capano}, \citenamefont {De}, \citenamefont {Cabero},
  \citenamefont {Brown}, \citenamefont {Nitz},\ and\ \citenamefont
  {Raymond}}]{Biwer2019}%
  \BibitemOpen
  \bibfield  {author} {\bibinfo {author} {\bibfnamefont {C.~M.}\ \bibnamefont
  {Biwer}}, \bibinfo {author} {\bibfnamefont {C.~D.}\ \bibnamefont {Capano}},
  \bibinfo {author} {\bibfnamefont {S.}~\bibnamefont {De}}, \bibinfo {author}
  {\bibfnamefont {M.}~\bibnamefont {Cabero}}, \bibinfo {author} {\bibfnamefont
  {D.~A.}\ \bibnamefont {Brown}}, \bibinfo {author} {\bibfnamefont {A.~H.}\
  \bibnamefont {Nitz}},\ and\ \bibinfo {author} {\bibfnamefont
  {V.}~\bibnamefont {Raymond}},\ }\bibfield  {title} {\bibinfo {title} {{PyCBC
  inference: a python-based parameter estimation toolkit for compact binary
  coalescence signals}},\ }\bibfield  {journal} {\bibinfo  {journal} {Publ.
  Astron. Soc. Pacific}\ }\textbf {\bibinfo {volume} {131}},\ \href
  {https://doi.org/10.1088/1538-3873/aaef0b} {10.1088/1538-3873/aaef0b}
  (\bibinfo {year} {2019}),\ \Eprint {https://arxiv.org/abs/1807.10312}
  {arXiv:1807.10312} \BibitemShut {NoStop}%
\bibitem [{\citenamefont {Ashton}\ \emph {et~al.}(2019)\citenamefont {Ashton},
  \citenamefont {Hübner}, \citenamefont {Lasky}, \citenamefont {Talbot},
  \citenamefont {Ackley}, \citenamefont {Biscoveanu}, \citenamefont {Chu},
  \citenamefont {Divakarla}, \citenamefont {Easter}, \citenamefont
  {Goncharov},\ and\ \citenamefont {et~al.}}]{Ashton2019}%
  \BibitemOpen
  \bibfield  {author} {\bibinfo {author} {\bibfnamefont {G.}~\bibnamefont
  {Ashton}}, \bibinfo {author} {\bibfnamefont {M.}~\bibnamefont {Hübner}},
  \bibinfo {author} {\bibfnamefont {P.~D.}\ \bibnamefont {Lasky}}, \bibinfo
  {author} {\bibfnamefont {C.}~\bibnamefont {Talbot}}, \bibinfo {author}
  {\bibfnamefont {K.}~\bibnamefont {Ackley}}, \bibinfo {author} {\bibfnamefont
  {S.}~\bibnamefont {Biscoveanu}}, \bibinfo {author} {\bibfnamefont
  {Q.}~\bibnamefont {Chu}}, \bibinfo {author} {\bibfnamefont {A.}~\bibnamefont
  {Divakarla}}, \bibinfo {author} {\bibfnamefont {P.~J.}\ \bibnamefont
  {Easter}}, \bibinfo {author} {\bibfnamefont {B.}~\bibnamefont {Goncharov}},\
  and\ \bibinfo {author} {\bibnamefont {et~al.}},\ }\bibfield  {title}
  {\bibinfo {title} {Bilby: A user-friendly bayesian inference library for
  gravitational-wave astronomy},\ }\href
  {https://doi.org/10.3847/1538-4365/ab06fc} {\bibfield  {journal} {\bibinfo
  {journal} {The Astrophysical Journal Supplement Series}\ }\textbf {\bibinfo
  {volume} {241}},\ \bibinfo {pages} {27} (\bibinfo {year} {2019})}\BibitemShut
  {NoStop}%
\bibitem [{\citenamefont {Buchner}\ \emph {et~al.}(2014)\citenamefont
  {Buchner}, \citenamefont {Georgakakis}, \citenamefont {Nandra}, \citenamefont
  {Hsu}, \citenamefont {Rangel}, \citenamefont {Brightman}, \citenamefont
  {Merloni}, \citenamefont {Salvato}, \citenamefont {Donley},\ and\
  \citenamefont {Kocevski}}]{PyMultinest}%
  \BibitemOpen
  \bibfield  {author} {\bibinfo {author} {\bibfnamefont {J.}~\bibnamefont
  {Buchner}}, \bibinfo {author} {\bibfnamefont {A.}~\bibnamefont
  {Georgakakis}}, \bibinfo {author} {\bibfnamefont {K.}~\bibnamefont {Nandra}},
  \bibinfo {author} {\bibfnamefont {L.}~\bibnamefont {Hsu}}, \bibinfo {author}
  {\bibfnamefont {C.}~\bibnamefont {Rangel}}, \bibinfo {author} {\bibfnamefont
  {M.}~\bibnamefont {Brightman}}, \bibinfo {author} {\bibfnamefont
  {A.}~\bibnamefont {Merloni}}, \bibinfo {author} {\bibfnamefont
  {M.}~\bibnamefont {Salvato}}, \bibinfo {author} {\bibfnamefont
  {J.}~\bibnamefont {Donley}},\ and\ \bibinfo {author} {\bibfnamefont
  {D.}~\bibnamefont {Kocevski}},\ }\bibfield  {title} {\bibinfo {title} {X-ray
  spectral modelling of the agn obscuring region in the cdfs: Bayesian model
  selection and catalogue},\ }\href
  {https://doi.org/10.1051/0004-6361/201322971} {\bibfield  {journal} {\bibinfo
   {journal} {Astronomy \& Astrophysics}\ }\textbf {\bibinfo {volume} {564}},\
  \bibinfo {pages} {A125} (\bibinfo {year} {2014})},\ \Eprint
  {https://arxiv.org/abs/1402.0004} {arXiv:1402.0004 [astro-ph.HE]}
  \BibitemShut {NoStop}%
\bibitem [{\citenamefont {London}\ \emph {et~al.}(2018)\citenamefont {London},
  \citenamefont {Khan}, \citenamefont {Fauchon-Jones}, \citenamefont
  {Garc{\'{i}}a}, \citenamefont {Hannam}, \citenamefont {Husa}, \citenamefont
  {Jim{\'{e}}nez-Forteza}, \citenamefont {Kalaghatgi}, \citenamefont {Ohme},\
  and\ \citenamefont {Pannarale}}]{London2018}%
  \BibitemOpen
  \bibfield  {author} {\bibinfo {author} {\bibfnamefont {L.}~\bibnamefont
  {London}}, \bibinfo {author} {\bibfnamefont {S.}~\bibnamefont {Khan}},
  \bibinfo {author} {\bibfnamefont {E.}~\bibnamefont {Fauchon-Jones}}, \bibinfo
  {author} {\bibfnamefont {C.}~\bibnamefont {Garc{\'{i}}a}}, \bibinfo {author}
  {\bibfnamefont {M.}~\bibnamefont {Hannam}}, \bibinfo {author} {\bibfnamefont
  {S.}~\bibnamefont {Husa}}, \bibinfo {author} {\bibfnamefont {X.}~\bibnamefont
  {Jim{\'{e}}nez-Forteza}}, \bibinfo {author} {\bibfnamefont {C.}~\bibnamefont
  {Kalaghatgi}}, \bibinfo {author} {\bibfnamefont {F.}~\bibnamefont {Ohme}},\
  and\ \bibinfo {author} {\bibfnamefont {F.}~\bibnamefont {Pannarale}},\
  }\bibfield  {title} {\bibinfo {title} {{First Higher-Multipole Model of
  Gravitational Waves from Spinning and Coalescing Black-Hole Binaries}},\
  }\href {https://doi.org/10.1103/PhysRevLett.120.161102} {\bibfield  {journal}
  {\bibinfo  {journal} {Phys. Rev. Lett.}\ }\textbf {\bibinfo {volume} {120}},\
  \bibinfo {pages} {2} (\bibinfo {year} {2018})},\ \Eprint
  {https://arxiv.org/abs/1708.00404} {arXiv:1708.00404} \BibitemShut {NoStop}%
\bibitem [{\citenamefont {Baibhav}\ \emph {et~al.}(2020)\citenamefont
  {Baibhav}, \citenamefont {Berti},\ and\ \citenamefont
  {Cardoso}}]{Baibhav_2020}%
  \BibitemOpen
  \bibfield  {author} {\bibinfo {author} {\bibfnamefont {V.}~\bibnamefont
  {Baibhav}}, \bibinfo {author} {\bibfnamefont {E.}~\bibnamefont {Berti}},\
  and\ \bibinfo {author} {\bibfnamefont {V.}~\bibnamefont {Cardoso}},\
  }\bibfield  {title} {\bibinfo {title} {Lisa parameter estimation and source
  localization with higher harmonics of the ringdown},\ }\href
  {https://doi.org/10.1103/physrevd.101.084053} {\bibfield  {journal} {\bibinfo
   {journal} {Physical Review D}\ }\textbf {\bibinfo {volume} {101}},\ \bibinfo
  {pages} {084053} (\bibinfo {year} {2020})}\BibitemShut {NoStop}%
\bibitem [{\citenamefont {García-Quirós}\ \emph {et~al.}(2020)\citenamefont
  {García-Quirós}, \citenamefont {Colleoni}, \citenamefont {Husa},
  \citenamefont {Estellés}, \citenamefont {Pratten}, \citenamefont
  {Ramos-Buades}, \citenamefont {Mateu-Lucena},\ and\ \citenamefont
  {Jaume}}]{IMRXHM2020}%
  \BibitemOpen
  \bibfield  {author} {\bibinfo {author} {\bibfnamefont {C.}~\bibnamefont
  {García-Quirós}}, \bibinfo {author} {\bibfnamefont {M.}~\bibnamefont
  {Colleoni}}, \bibinfo {author} {\bibfnamefont {S.}~\bibnamefont {Husa}},
  \bibinfo {author} {\bibfnamefont {H.}~\bibnamefont {Estellés}}, \bibinfo
  {author} {\bibfnamefont {G.}~\bibnamefont {Pratten}}, \bibinfo {author}
  {\bibfnamefont {A.}~\bibnamefont {Ramos-Buades}}, \bibinfo {author}
  {\bibfnamefont {M.}~\bibnamefont {Mateu-Lucena}},\ and\ \bibinfo {author}
  {\bibfnamefont {R.}~\bibnamefont {Jaume}},\ }\bibfield  {title} {\bibinfo
  {title} {Multimode frequency-domain model for the gravitational wave signal
  from nonprecessing black-hole binaries},\ }\href
  {https://doi.org/10.1103/physrevd.102.064002} {\bibfield  {journal} {\bibinfo
   {journal} {Physical Review D}\ }\textbf {\bibinfo {volume} {102}},\ \bibinfo
  {pages} {064002} (\bibinfo {year} {2020})}\BibitemShut {NoStop}%
\bibitem [{\citenamefont {Blanchet}(2014)}]{Blanchet2014}%
  \BibitemOpen
  \bibfield  {author} {\bibinfo {author} {\bibfnamefont {L.}~\bibnamefont
  {Blanchet}},\ }\bibfield  {title} {\bibinfo {title} {{Gravitational radiation
  from post-newtonian sources and inspiralling compact binaries}},\ }\href
  {https://doi.org/10.12942/lrr-2014-2} {\bibfield  {journal} {\bibinfo
  {journal} {Living Rev. Relativ.}\ }\textbf {\bibinfo {volume} {17}},\
  \bibinfo {pages} {1} (\bibinfo {year} {2014})},\ \Eprint
  {https://arxiv.org/abs/1310.1528} {arXiv:1310.1528} \BibitemShut {NoStop}%
\bibitem [{\citenamefont {Farr}\ \emph {et~al.}(2016)\citenamefont {Farr},
  \citenamefont {Berry}, \citenamefont {Farr}, \citenamefont {Haster},
  \citenamefont {Middleton}, \citenamefont {Cannon}, \citenamefont {Graff},
  \citenamefont {Hanna}, \citenamefont {Mandel}, \citenamefont {Pankow},\ and\
  \citenamefont {et~al.}}]{SpinParaEst}%
  \BibitemOpen
  \bibfield  {author} {\bibinfo {author} {\bibfnamefont {B.}~\bibnamefont
  {Farr}}, \bibinfo {author} {\bibfnamefont {C.~P.~L.}\ \bibnamefont {Berry}},
  \bibinfo {author} {\bibfnamefont {W.~M.}\ \bibnamefont {Farr}}, \bibinfo
  {author} {\bibfnamefont {C.-J.}\ \bibnamefont {Haster}}, \bibinfo {author}
  {\bibfnamefont {H.}~\bibnamefont {Middleton}}, \bibinfo {author}
  {\bibfnamefont {K.}~\bibnamefont {Cannon}}, \bibinfo {author} {\bibfnamefont
  {P.~B.}\ \bibnamefont {Graff}}, \bibinfo {author} {\bibfnamefont
  {C.}~\bibnamefont {Hanna}}, \bibinfo {author} {\bibfnamefont
  {I.}~\bibnamefont {Mandel}}, \bibinfo {author} {\bibfnamefont
  {C.}~\bibnamefont {Pankow}},\ and\ \bibinfo {author} {\bibnamefont
  {et~al.}},\ }\bibfield  {title} {\bibinfo {title} {Parameter estimation on
  gravitational waves from neutron-star binaries with spinning components},\
  }\href {https://doi.org/10.3847/0004-637x/825/2/116} {\bibfield  {journal}
  {\bibinfo  {journal} {The Astrophysical Journal}\ }\textbf {\bibinfo {volume}
  {825}},\ \bibinfo {pages} {116} (\bibinfo {year} {2016})},\ \Eprint
  {https://arxiv.org/abs/1508.05336} {arXiv:1508.05336} \BibitemShut {NoStop}%
\bibitem [{\citenamefont {Zhao}\ \emph
  {et~al.}(2020{\natexlab{b}})\citenamefont {Zhao}, \citenamefont {Liu},
  \citenamefont {Wen}, \citenamefont {Zhu}, \citenamefont {Wang}, \citenamefont
  {Hu},\ and\ \citenamefont {Zhou}}]{Zhao2019}%
  \BibitemOpen
  \bibfield  {author} {\bibinfo {author} {\bibfnamefont {W.}~\bibnamefont
  {Zhao}}, \bibinfo {author} {\bibfnamefont {T.}~\bibnamefont {Liu}}, \bibinfo
  {author} {\bibfnamefont {L.}~\bibnamefont {Wen}}, \bibinfo {author}
  {\bibfnamefont {T.}~\bibnamefont {Zhu}}, \bibinfo {author} {\bibfnamefont
  {A.}~\bibnamefont {Wang}}, \bibinfo {author} {\bibfnamefont {Q.}~\bibnamefont
  {Hu}},\ and\ \bibinfo {author} {\bibfnamefont {C.}~\bibnamefont {Zhou}},\
  }\bibfield  {title} {\bibinfo {title} {Model-independent test of the parity
  symmetry of gravity with gravitational waves},\ }\href
  {https://doi.org/10.1140/epjc/s10052-020-8211-4} {\bibfield  {journal}
  {\bibinfo  {journal} {The European Physical Journal C}\ }\textbf {\bibinfo
  {volume} {80}},\ \bibinfo {pages} {630} (\bibinfo {year}
  {2020}{\natexlab{b}})}\BibitemShut {NoStop}%
\bibitem [{\citenamefont {Okounkova}\ \emph {et~al.}(2021)\citenamefont
  {Okounkova}, \citenamefont {Farr}, \citenamefont {Isi},\ and\ \citenamefont
  {Stein}}]{okounkova2021constraining}%
  \BibitemOpen
  \bibfield  {author} {\bibinfo {author} {\bibfnamefont {M.}~\bibnamefont
  {Okounkova}}, \bibinfo {author} {\bibfnamefont {W.~M.}\ \bibnamefont {Farr}},
  \bibinfo {author} {\bibfnamefont {M.}~\bibnamefont {Isi}},\ and\ \bibinfo
  {author} {\bibfnamefont {L.~C.}\ \bibnamefont {Stein}},\ }\href@noop {}
  {\bibinfo {title} {Constraining gravitational wave amplitude birefringence
  and chern-simons gravity with gwtc-2}} (\bibinfo {year} {2021}),\ \Eprint
  {https://arxiv.org/abs/2101.11153} {arXiv:2101.11153 [gr-qc]} \BibitemShut
  {NoStop}%
\bibitem [{\citenamefont {{Abbott}}\ \emph {et~al.}(2017)\citenamefont
  {{Abbott}}, \citenamefont {{Abbott}}, \citenamefont {{Abbott}}, \citenamefont
  {{Acernese}}, \citenamefont {{Ackley}}, \citenamefont {{Adams}},
  \citenamefont {{Adams}}, \citenamefont {{Addesso}}, \citenamefont
  {{Adhikari}}, \citenamefont {{Adya}},\ and\ \citenamefont
  {et~al.}}]{GWspeed1}%
  \BibitemOpen
  \bibfield  {author} {\bibinfo {author} {\bibfnamefont {B.~P.}\ \bibnamefont
  {{Abbott}}}, \bibinfo {author} {\bibfnamefont {R.}~\bibnamefont {{Abbott}}},
  \bibinfo {author} {\bibfnamefont {T.~D.}\ \bibnamefont {{Abbott}}}, \bibinfo
  {author} {\bibfnamefont {F.}~\bibnamefont {{Acernese}}}, \bibinfo {author}
  {\bibfnamefont {K.}~\bibnamefont {{Ackley}}}, \bibinfo {author}
  {\bibfnamefont {C.}~\bibnamefont {{Adams}}}, \bibinfo {author} {\bibfnamefont
  {T.}~\bibnamefont {{Adams}}}, \bibinfo {author} {\bibfnamefont
  {P.}~\bibnamefont {{Addesso}}}, \bibinfo {author} {\bibfnamefont {R.~X.}\
  \bibnamefont {{Adhikari}}}, \bibinfo {author} {\bibfnamefont {V.~B.}\
  \bibnamefont {{Adya}}},\ and\ \bibinfo {author} {\bibnamefont {et~al.}},\
  }\bibfield  {title} {\bibinfo {title} {{Gravitational Waves and Gamma-Rays
  from a Binary Neutron Star Merger: GW170817 and GRB 170817A}},\ }\href
  {https://doi.org/10.3847/2041-8213/aa920c} {\bibfield  {journal} {\bibinfo
  {journal} {Astrophys. J. Lett.}\ }\textbf {\bibinfo {volume} {848}},\
  \bibinfo {eid} {L13} (\bibinfo {year} {2017})},\ \Eprint
  {https://arxiv.org/abs/1710.05834} {arXiv:1710.05834 [astro-ph.HE]}
  \BibitemShut {NoStop}%
\bibitem [{\citenamefont {Nishizawa}\ and\ \citenamefont
  {Kobayashi}(2018)}]{GWspeed2}%
  \BibitemOpen
  \bibfield  {author} {\bibinfo {author} {\bibfnamefont {A.}~\bibnamefont
  {Nishizawa}}\ and\ \bibinfo {author} {\bibfnamefont {T.}~\bibnamefont
  {Kobayashi}},\ }\bibfield  {title} {\bibinfo {title} {Parity-violating
  gravity and gq170817},\ }\href {https://doi.org/10.1103/PhysRevD.98.124018}
  {\bibfield  {journal} {\bibinfo  {journal} {Phys. Rev. D}\ }\textbf {\bibinfo
  {volume} {98}},\ \bibinfo {pages} {124018} (\bibinfo {year}
  {2018})}\BibitemShut {NoStop}%
\bibitem [{\citenamefont {Smith}\ \emph {et~al.}(2008)\citenamefont {Smith},
  \citenamefont {Erickcek}, \citenamefont {Caldwell},\ and\ \citenamefont
  {Kamionkowski}}]{solarsystem}%
  \BibitemOpen
  \bibfield  {author} {\bibinfo {author} {\bibfnamefont {T.~L.}\ \bibnamefont
  {Smith}}, \bibinfo {author} {\bibfnamefont {A.~L.}\ \bibnamefont {Erickcek}},
  \bibinfo {author} {\bibfnamefont {R.~R.}\ \bibnamefont {Caldwell}},\ and\
  \bibinfo {author} {\bibfnamefont {M.}~\bibnamefont {Kamionkowski}},\
  }\bibfield  {title} {\bibinfo {title} {Effects of chern-simons gravity on
  bodies orbiting the earth},\ }\href
  {https://doi.org/10.1103/PhysRevD.77.024015} {\bibfield  {journal} {\bibinfo
  {journal} {Phys. Rev. D}\ }\textbf {\bibinfo {volume} {77}},\ \bibinfo
  {pages} {024015} (\bibinfo {year} {2008})}\BibitemShut {NoStop}%
\bibitem [{\citenamefont {Yunes}\ and\ \citenamefont
  {Spergel}(2009)}]{binarypulsar1}%
  \BibitemOpen
  \bibfield  {author} {\bibinfo {author} {\bibfnamefont {N.}~\bibnamefont
  {Yunes}}\ and\ \bibinfo {author} {\bibfnamefont {D.~N.}\ \bibnamefont
  {Spergel}},\ }\bibfield  {title} {\bibinfo {title} {Double-binary-pulsar test
  of chern-simons modified gravity},\ }\href
  {https://doi.org/10.1103/PhysRevD.80.042004} {\bibfield  {journal} {\bibinfo
  {journal} {Phys. Rev. D}\ }\textbf {\bibinfo {volume} {80}},\ \bibinfo
  {pages} {042004} (\bibinfo {year} {2009})}\BibitemShut {NoStop}%
\bibitem [{\citenamefont {Ali-Haimoud}(2011)}]{binarypulsar2}%
  \BibitemOpen
  \bibfield  {author} {\bibinfo {author} {\bibfnamefont {Y.}~\bibnamefont
  {Ali-Haimoud}},\ }\bibfield  {title} {\bibinfo {title} {Revisiting the
  double-binary-pulsar probe of nondynamical chern-simons gravity},\ }\href
  {https://doi.org/10.1103/PhysRevD.83.124050} {\bibfield  {journal} {\bibinfo
  {journal} {Phys. Rev. D}\ }\textbf {\bibinfo {volume} {83}},\ \bibinfo
  {pages} {124050} (\bibinfo {year} {2011})}\BibitemShut {NoStop}%
\bibitem [{\citenamefont {{Arun}}\ \emph {et~al.}(2004)\citenamefont {{Arun}},
  \citenamefont {{Blanchet}}, \citenamefont {{Iyer}},\ and\ \citenamefont
  {{Qusailah}}}]{2004CQGra213771A}%
  \BibitemOpen
  \bibfield  {author} {\bibinfo {author} {\bibfnamefont {K.~G.}\ \bibnamefont
  {{Arun}}}, \bibinfo {author} {\bibfnamefont {L.}~\bibnamefont {{Blanchet}}},
  \bibinfo {author} {\bibfnamefont {B.~R.}\ \bibnamefont {{Iyer}}},\ and\
  \bibinfo {author} {\bibfnamefont {M.~S.~S.}\ \bibnamefont {{Qusailah}}},\
  }\bibfield  {title} {\bibinfo {title} {{The 2.5PN gravitational wave
  polarizations from inspiralling compact binaries in circular orbits}},\
  }\href {https://doi.org/10.1088/0264-9381/21/15/010} {\bibfield  {journal}
  {\bibinfo  {journal} {Classical and Quantum Gravity}\ }\textbf {\bibinfo
  {volume} {21}},\ \bibinfo {pages} {3771} (\bibinfo {year} {2004})},\ \Eprint
  {https://arxiv.org/abs/gr-qc/0404085} {arXiv:gr-qc/0404085 [gr-qc]}
  \BibitemShut {NoStop}%
\bibitem [{\citenamefont {Arun}\ \emph {et~al.}(2005)\citenamefont {Arun},
  \citenamefont {Blanchet}, \citenamefont {Iyer},\ and\ \citenamefont
  {Qusailah}}]{Arun_2005}%
  \BibitemOpen
  \bibfield  {author} {\bibinfo {author} {\bibfnamefont {K.~G.}\ \bibnamefont
  {Arun}}, \bibinfo {author} {\bibfnamefont {L.}~\bibnamefont {Blanchet}},
  \bibinfo {author} {\bibfnamefont {B.~R.}\ \bibnamefont {Iyer}},\ and\
  \bibinfo {author} {\bibfnamefont {M.~S.~S.}\ \bibnamefont {Qusailah}},\
  }\bibfield  {title} {\bibinfo {title} {The 2.5pn gravitational wave
  polarizations from inspiralling compact binaries in circular orbits},\ }\href
  {https://doi.org/10.1088/0264-9381/22/14/c01} {\bibfield  {journal} {\bibinfo
   {journal} {Classical and Quantum Gravity}\ }\textbf {\bibinfo {volume}
  {22}},\ \bibinfo {pages} {3115} (\bibinfo {year} {2005})}\BibitemShut
  {NoStop}%
\bibitem [{\citenamefont {{Van Den Broeck}}\ and\ \citenamefont
  {{Sengupta}}(2007)}]{2007CQGra24155V}%
  \BibitemOpen
  \bibfield  {author} {\bibinfo {author} {\bibfnamefont {C.}~\bibnamefont {{Van
  Den Broeck}}}\ and\ \bibinfo {author} {\bibfnamefont {A.~S.}\ \bibnamefont
  {{Sengupta}}},\ }\bibfield  {title} {\bibinfo {title} {{Phenomenology of
  amplitude-corrected post-Newtonian gravitational waveforms for compact binary
  inspiral: I. Signal-to-noise ratios}},\ }\href
  {https://doi.org/10.1088/0264-9381/24/1/009} {\bibfield  {journal} {\bibinfo
  {journal} {Classical and Quantum Gravity}\ }\textbf {\bibinfo {volume}
  {24}},\ \bibinfo {pages} {155} (\bibinfo {year} {2007})},\ \Eprint
  {https://arxiv.org/abs/gr-qc/0607092} {arXiv:gr-qc/0607092 [gr-qc]}
  \BibitemShut {NoStop}%
\bibitem [{\citenamefont {{Zhao}}\ and\ \citenamefont
  {{Wen}}(2018)}]{zhao2018}%
  \BibitemOpen
  \bibfield  {author} {\bibinfo {author} {\bibfnamefont {W.}~\bibnamefont
  {{Zhao}}}\ and\ \bibinfo {author} {\bibfnamefont {L.}~\bibnamefont {{Wen}}},\
  }\bibfield  {title} {\bibinfo {title} {{Localization accuracy of compact
  binary coalescences detected by the third-generation gravitational-wave
  detectors and implication for cosmology}},\ }\href
  {https://doi.org/10.1103/PhysRevD.97.064031} {\bibfield  {journal} {\bibinfo
  {journal} {\prd}\ }\textbf {\bibinfo {volume} {97}},\ \bibinfo {eid} {064031}
  (\bibinfo {year} {2018})},\ \Eprint {https://arxiv.org/abs/1710.05325}
  {arXiv:1710.05325 [astro-ph.CO]} \BibitemShut {NoStop}%
\bibitem [{\citenamefont {Niu}\ \emph {et~al.}(2020)\citenamefont {Niu},
  \citenamefont {Zhang}, \citenamefont {Liu}, \citenamefont {Yu}, \citenamefont
  {Wang},\ and\ \citenamefont {Zhao}}]{Niu2020}%
  \BibitemOpen
  \bibfield  {author} {\bibinfo {author} {\bibfnamefont {R.}~\bibnamefont
  {Niu}}, \bibinfo {author} {\bibfnamefont {X.}~\bibnamefont {Zhang}}, \bibinfo
  {author} {\bibfnamefont {T.}~\bibnamefont {Liu}}, \bibinfo {author}
  {\bibfnamefont {J.}~\bibnamefont {Yu}}, \bibinfo {author} {\bibfnamefont
  {B.}~\bibnamefont {Wang}},\ and\ \bibinfo {author} {\bibfnamefont
  {W.}~\bibnamefont {Zhao}},\ }\bibfield  {title} {\bibinfo {title}
  {{Constraining Screened Modified Gravity with Spaceborne Gravitational-wave
  Detectors}},\ }\href {https://doi.org/10.3847/1538-4357/ab6d03} {\bibfield
  {journal} {\bibinfo  {journal} {Astrophys. J.}\ }\textbf {\bibinfo {volume}
  {890}},\ \bibinfo {pages} {163} (\bibinfo {year} {2020})},\ \Eprint
  {https://arxiv.org/abs/1910.10592} {arXiv:1910.10592} \BibitemShut {NoStop}%
\end{thebibliography}%

\end{document}